\documentclass[a4paper,10pt]{paper}

\usepackage{fullpage}
\usepackage{parskip}
\usepackage{amssymb,amsmath}

\usepackage{graphicx}
\usepackage{fixltx2e}
\usepackage{grffile}
\usepackage[T1]{fontenc}
\usepackage{xcolor}
\usepackage{subfig}
\usepackage{marvosym}
\usepackage{fancyvrb}
\usepackage{tabularx}
\usepackage{booktabs}

\usepackage{natbib}
\usepackage[pdftex]{hyperref}

\hypersetup{colorlinks=true, bookmarksopen=true, bookmarksnumbered=true,pdfstartview=FitV, linktocpage = true, bookmarksopen = true,
    citecolor=blue, filecolor=blue, linkcolor=blue, urlcolor=blue}

\usepackage{subfig}

\renewcommand{\vec}[1]{\mbox{\boldmath{$#1$}}}
\newcommand{\mat}[1]{\mbox{$\mathrm{#1}$}}
\newcommand{\pipe}{\; \middle\vert  \;}

\usepackage{authblk}

\let\oldmarginpar\marginpar
\renewcommand\marginpar[1]{\-\oldmarginpar[\raggedleft\footnotesize \sffamily \color{gray} #1]%
{\raggedright\footnotesize \sffamily \color{gray} #1}}

\usepackage{mdwlist}

\begin{document}

\title{A Bayesian spatio-temporal model of panel design data: airborne particle number concentration in Brisbane, Australia}

\author[1,2]{Sam Clifford}
\affil[1]{International Laboratory for Air Quality and Health, Queensland University of Technology, GPO Box 2434, Brisbane, Qld 4000, Australia}
\affil[2]{Centre for Air Quality \& Health Research and Evaluation, 431 Glebe Point Rd, Glebe 2037, Australia}
\author[3]{Sama {Low Choy}}
\affil[3]{School of Mathematical Sciences, Science and Engineering Faculty, Queensland University of Technology, GPO Box 2434, Brisbane, Qld 4000, Australia}
\author[1]{Mandana Mazaheri}
\author[1]{Farhad Salimi}
\author[1]{Lidia Morawska}
\author[3,\Email]{Kerrie Mengersen}
\affil[\Email]{Corresponding author: \href{mailto:k.mengersen@qut.edu.au}{k.mengersen@qut.edu.au}, ph +61 7 3138 2063}

\maketitle


\begin{abstract}
In environmental monitoring, the ability to obtain high quality data across space and time is often limited by the cost of purchasing, deploying and maintaining a large collection of equipment and the employment of personnel to perform these tasks. An ideal design for a monitoring campaign would be dense enough in time to capture short-range variation at each site, long enough in time to examine trends at each site and across all sites, and dense enough in space to allow modelling of the relationship between the means at each of the sites.

This paper outlines a methodology for semi-parametric spatio-temporal modelling of data which is dense in time but sparse in space, obtained from a split panel design, the most feasible approach to covering space and time with limited equipment. The data are hourly averaged particle number concentration (PNC) and were collected, as part of the International Laboratory for Air Quality and Health's Ultrafine Particles from Transport Emissions and Child Health (UPTECH) project. Two weeks of continuous measurements were taken at each of a number of government primary schools in the Brisbane Metropolitan Area. The monitoring equipment was taken to each school sequentially. The school data are augmented by data from long term monitoring stations at three locations in Brisbane, Australia.

Fitting the model helps describe the spatial and temporal variability at a subset of the UPTECH schools and the long-term monitoring sites. The temporal variation is modelled hierarchically with a penalised random walk term common to all sites and a similar term accounting for the remaining temporal trend at each site. The modelling of temporal trends requires an acknowledgement that the observations are correlated rather than independent. Parameter estimates and their uncertainty are computed in a computationally efficient approximate Bayesian inference environment, R-INLA.

The temporal part of the model explains daily and weekly cycles in PNC at the schools, which can be used to estimate the exposure of school children to ultrafine particles (UFPs) emitted by vehicles. At each school and long-term monitoring site, peaks in PNC can be attributed to the morning and afternoon rush hour traffic and new particle formation events. The spatial component of the model describes the school to school variation in mean PNC at each school and within each school ground. It is shown how the spatial model can be expanded to identify spatial patterns at the city scale with the inclusion of more spatial locations.
\end{abstract}


\section{Introduction}
Panel designs are a class of experimental study design where the aim is to understand spatial and temporal variation but continuous and contemporaneous measurement at all sites is not feasible \citep{dobbie08}. The number of sites and length of the measurement campaign at each site may be limited by such factors as cost of equipment and availability of trained staff to deploy, operate and maintain the equipment. The competing goals of quantifying spatial variation with short measurement campaigns at many sites and temporal variation with long campaigns at few sites can be achieved with the split panel design. The split panel design comprises a small number of locations where measurements are continuous and many locations where measurements are obtained for a short time before moving on to the next site.

The data were collected as part of the ``Ultrafine Particles from Transport Emissions and Child Health'' (UPTECH) project\footnote{\url{http://www.ilaqh.qut.edu.au/Misc/UPTECH\%20Home.htm}}. The main hypothesis of the UPTECH project is that UFPs affect the long term health of school children aged 8-11 years. As the first cohort study of the health effects of UFPs on children, UPTECH has characterised long-term health in terms of  respiratory, inflammatory and endothelial (pertaining to the cell wall of the lymphatic and blood vessels) health outcomes. For each child, exposure to traffic sources of UFPs at school was considered to be a large and measurable component of exposure.

The school environment has been found to be a major contributor to exposure to air pollutants such as nitrous oxides, sulphur dioxide, ozone and PM$_{10}$ \citep{UPTECH2010}. By number, a majority of the particles in outdoor, ambient air in Brisbane are in the ultrafine size range (diameter less than 100nm) and are generated by traffic \citep{characterisation98}. While it is known that exposure to traffic-generated air pollution has negative health effects \citet{hei2010} there has been no major study into the health effects of ultrafine particles on children within the schooling environment. The exposure of children to ultrafine particles from traffic can thus be quantified by measuring PNC at schools and at long term monitoring stations in the Brisbane Metropolitan Area. PNC can then be estimated as a function of space and time and the exposure estimated by integrating this spatio-temporal function along the paths in space-time along which school children travel.

The aim of this paper is to devise a statistical model for estimating spatio-temporal variation in ultrafine PNC over the spatial locations and temporal range of the UPTECH project. This is achieved by developing a hierarchical regression model for data measured according to a split panel design. The modelling goals are to describe the regional temporal trends common to all sites, the site-specific temporal trends and the mean level of PNC in primary schools in the Brisbane Metropolitan Area. The developed model will be used in future work to quantify school childrens' exposure to ultrafine particles.

To achieve the aim of this paper, the following questions will be addressed. In terms of the spatial modelling:
\begin{enumerate}
    \item Is the background level the same across all sites? \label{aim:sites}
    \item Is there a difference in background level for each location within each site? \label{aim:loc}
\suspend{enumerate}

Regarding quantification of the daily and weekly trends at each school and long-term monitoring station in the split panel design:
\resume{enumerate}
    \item Do the trends differ from site to site? \label{aim:trend}
    \item Can we avoid enforcing a particular form of the trends but still obtain smooth estimates that sensibly describe the temporal variation? \label{aim:np}
\suspend{enumerate}

In terms of the modelling approach, a spatio-temporal model for split panel design data fit in an approximate inference package:
\resume{enumerate}
    \item What is gained by modelling the trends at each site hierarchically? \label{aim:hier}
    \item How can we ensure convergence in such a complex model? \label{aim:conv}
\end{enumerate}

The model was developed and fit in R-INLA \citep{inlapackage}, an R package for approximate Bayesian inference. The R-INLA package uses the Integrated, Nested Laplace Approximation (INLA) to represent the terms in a latent Gaussian model (a general class of models which includes the Generalised Linear Model, Generalised Additive Model, Generalised Linear Mixed Model, and Generalised Additive Mixed Model and their extensions) as a Gaussian Markov Random Field (GMRF) \citep{inla-jrssb09}. The R-INLA approach allows for fast computation of complex models as it incorporates a Newton solver to find the maximum of the posterior density rather than sampling as in MCMC methods.

The model developed in this paper develops custom prior precision matrices for the GMRFs representations of latent Gaussian models. The custom precision matrices enforce smoothness in the estimates of the cyclic daily-weekly trend with a smoothness penalty matrix, rather than in the likelihood. The smoothness penalty approach minimises the difference between observations an hour apart and those a day apart without requiring the specification of a non-parametric spline basis or parametric Fourier series basis. This modelling approach allows the borrowing of strength across the small number of observations at each of the short-term monitoring locations in the split panel design by focussing on the relationship between observations rather than basis elements.

The spatial variation is modelled with a spatial random effect based on a stochastic partial differential equation (SPDE) representation of the GMRF approximation to a Gaussian process \citep{lindgrenJRSSB}. The SPDE approach involves discretising the study domain with an unstructured triangular mesh with a finite element basis. The solution to the SPDE is a GMRF with a Mat\'ern class covariance function.

The resulting model is a complex, sparse GMRF which includes terms for the hierarchical temporal trends and a sparse spatial random effect.

\section{Data}\label{sec:data}
\begin{figure}[ht]
\centering
\includegraphics[width=0.6\linewidth]{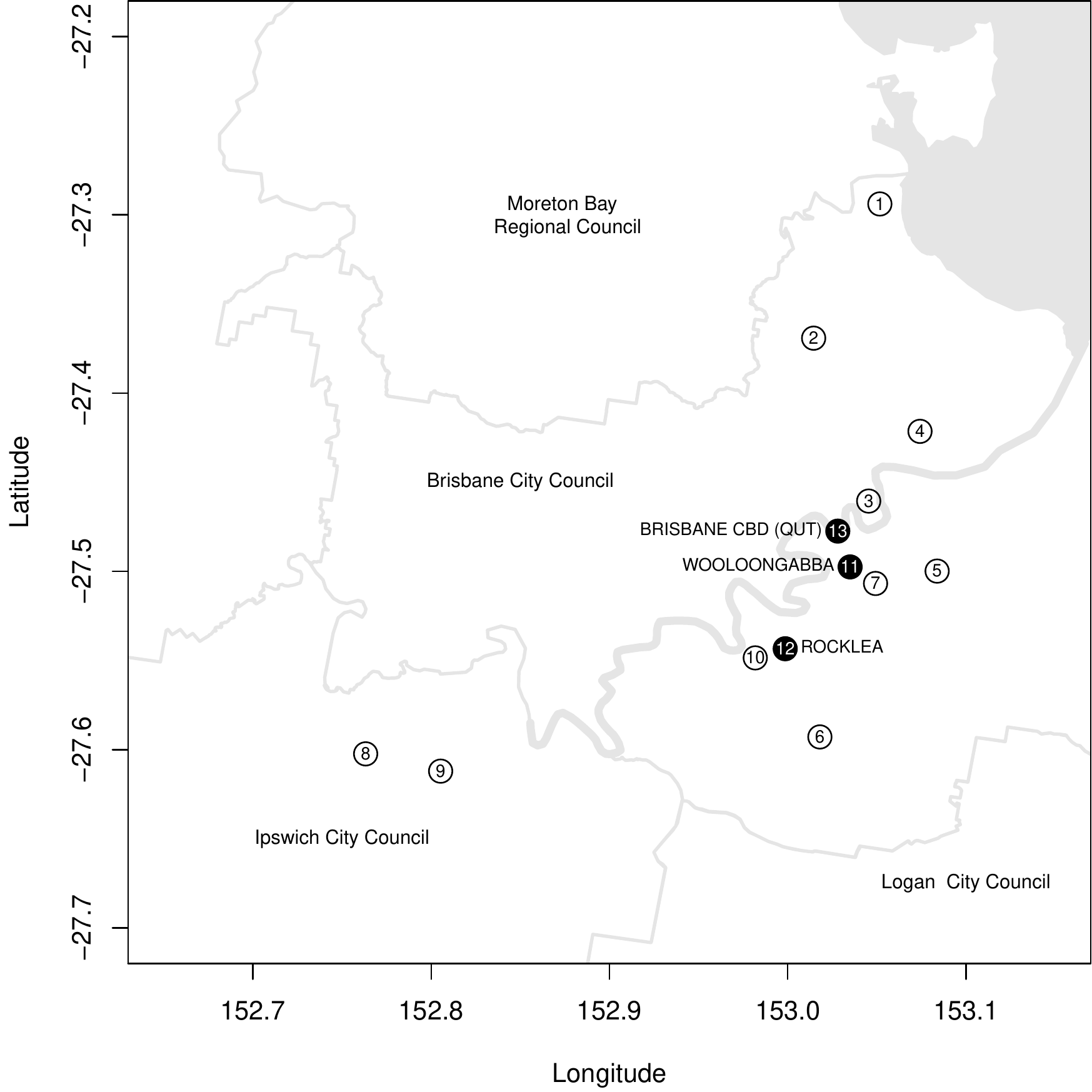}
\caption{Schools 1 to 10 and long term air quality monitoring stations.}
\label{fig:tenschools}
\end{figure}

The UPTECH study involves the short-term measurement of PNC at each of 25 of primary schools in the Brisbane Metropolitan Area and three long-term monitoring stations. To perform the spatio-temporal modelling in this paper, data from the first ten schools are used (Figure \ref{fig:tenschools}), which gives an adequate number of sites to illustrate the set up and use of the model prior to the completion of the validation and collation of all data from the project. Measurement of PNC in the UPTECH project was subject to a split panel design, where the two week measurement campaigns at each school were augmented by long term monitoring (Table \ref{tab:schools}). 

\begin{table}[ht]
\centering
{\footnotesize
\begin{tabularx}{0.8\linewidth}{rlXrr}
\toprule
Site & Prevailing winds & Features & Start & End \\
\midrule
1     & E to NE & Arterial road & 15/11/10 & 28/11/10 \\
2     & E & Arterial road & 18/10/10 & 31/10/10 \\
3     & NE to E & River & 1/11/10 & 14/11/10 \\
4     & All but W to NW & Industrial and airport & 28/2/11 & 13/3/11 \\
5     & SW to SE & Arterial road & 21/3/11 & 3/4/11 \\
6     & S to W & Semi-rural, arterial road & 16/5/11 & 29/5/11 \\
7     & S to W & Elevated freeway & 30/5/11 & 12/6/11 \\
8     & SW to W & Industrial & 14/6/11 & 24/6/11 \\
9     & SW to W & Open spaces and train line & 11/7/11 & 24/7/11 \\
10    & S to SW & Train lines and arterial road & 25/7/11 & 7/8/11 \\
11 & S to SE & Arterial road & 1/1/09 & 31/8/10 \\ 
12 & S to W, NE & Industrial & 1/1/09 & 31/8/10 \\
13 & SW, E & CBD, freeway & 1/1/09 &  31/12/09 \\
 &  &  & 20/9/10 & 16/8/11 \\
\bottomrule
\end{tabularx}
}
\caption{Description of each measurement site. Sites 1 to 10 are the first ten schools in the UPTECH project. The remaining sites are Woolloongabba -- 11, Rocklea -- 12, Queensland University of Technology -- 13.}
\label{tab:schools}
\end{table}

An ideal split panel design has measurements being taken at the long term monitoring stations throughout the entire project length and no breaks between the monitoring at the short term sites. Due to the timing of school holidays and the cessation of PNC recording at Rocklea and Woolloongabba by DERM, the actual duration of measurements differs from this ideal design. The monitoring at QUT continued for the duration of the UPTECH campaign.

PNC was recorded at each site with Condensation Particle Counters (CPCs), devices which saturate a sample of air with a vapour (commonly water or butanol) and then count the number of droplets with a light scattering technique \citep{cnc93}. The specific CPC instruments used for measurement were the TSI 3781 and 3787. The model 3781 can detect particles in the range 6 to 3000 nm; the 3787 can detect particles from 5 nm to 3000 nm. Three CPCs were deployed at each school in order to characterise school-scale spatial variation. The CPCs at the schools were labelled ``A'', ``B'' and ``C'' based on their position. As often as possible, CPC ``A'' was located such that it was immediately downwind of the largest road near the school in order to capture the highest PNC that was likely present at the school. CPC ``C'' was located far away from ``A'' in the downwind direction to capture the background levels at the school, away from the largest road. CPC ``B'' was located in the middle of the school but not necessarily colinear with the other two CPCs. The ``downwind'' direction was determined by examining prevailing winds at the nearest Bureau of Meteorology weather station, averaging the wind speed and direction information for the planned month of measurement at that school, going back to 2000.

Schools in the UPTECH study were labelled sequentially based on the date of the measurement at that school (with the exception of the first three schools, which were measured in the order 2, 3, 1). The first ten schools, therefore, are labelled here 1 to 10 and the monitoring stations are labelled as follows: Woolloongabba (11), Rocklea (12) and QUT (13). The measurements of PNC from the split panel design feature 20 months of continuous monitoring at both Woolloongabba and Rocklea. The measurements at QUT are one calendar year of continuous monitoring and 11 months beginning shortly after the cessation of monitoring of PNC at Woolloongabba and Rocklea. The second round of monitoring at QUT was concurrent with the school-based measurements at schools 1 to 10 and continued throughout the UPTECH project.

The application of this research is the modelling of the spatio-temporal distribution of ultrafine particles and, ultimately, the estimation of the exposure of school children to these particles. The cost of monitoring ultrafine particles at multiple locations motivates the use of a split panel design and the development of a spatio-temporal model for data collected from such a design allows the estimation of exposure to ultrafine particles.

\section{Model}\label{sec:meth}
It has been shown that PNC in Brisbane exhibits daily and weekly trends \citep{weeklytrends2002,jaime5year} and accounting for these trends will form the basis of the temporal aspect of the model. PNC in Brisbane is also spatially variable \citep{jaimespatial} and any model which features measurement at multiple locations should account for possible spatial variation.

The regression model fit to the data, outlined below, is a Bayesian semi-parametric additive model with a Normal likelihood for the log of particle number concentration. Temporal trends common to all sites and specific to each site will be fit. Spatial variation in log PNC will be modelled with spatial random effect for the mean at each measurement location. These model components will be represented as latent Gaussian models using Gaussian Markov Random Fields. The hierarchical nature of the model, imposed by the panel design, arises through the fitting of an all-sites mean, temporal trends common to all sites and specific to each site and a spatially varying random mean,
\begin{equation}
\begin{aligned}
 \log y_{ij} =& \beta_0 + \beta_{1j} + f_j\left( \textnormal{hour}_{ij}, \textnormal{day of week}_{ij} \right) + f_{\textnormal{year}}\left( \textnormal{day of year} \right)_i + \varepsilon_{ij} \\
 \varepsilon_{ij} \sim & \mathcal{N}\left(0, \sigma^2 \right)
\end{aligned} \label{eq:model}
\end{equation}
where $f_j(\cdot,\cdot)$ represents the temporal trends at site $j$ and may be additive functions of the hour of the day and day of the week or some joint, non-separable function to be estimated. In a similar manner, $f_{\textnormal{year}}\left( \textnormal{day of year} \right)_i$ is an annual trend which is to be estimated via non-parametric techniques described below. The annual trend is required to deal with the confounding of the spatial and temporal variation due to the split panel design. Here $\beta_0$ is the overall mean across all sites, with an uninformative normal prior, and $\beta_{1j}$ is the mean at site $j$, modelled with the spatial random effect described in Section \ref{sec:spde}.

Temporal trends in time series data may be modelled with non-parametric scatterplot smoothers such as splines \citep{harveykoopman93}, local polynomial regression \citep{loess} or random walk smoothers \citep{schrodleJRSSC11}. The scatterplot smoothers are able to adapt to local changes in covariate behaviour and thus model non-linear effects without relying on a particular parametric formulation of the basis \citep{hastietib}.

The periodicity in these data can be modelled with, among others, splines with periodic basis functions \citep{harveykoopman93} or with a random walk model with a Toeplitz circulant penalty matrix \citep{gmrfbook}. Traditional time series models, such as models based on the approach of \citet*{boxjenkins}, aim to quantify the relationship between observations rather than estimate smooth temporal trends. These models express the temporal relationship as a linear combination of previous observations rather than as a smooth trend. Seasonal decomposition by loess \citep{stl} characterises seasonality in the data with local polynomial regression at a coarse time scale and estimates a cyclic trend with a smaller span local polynomial regression but the smoothness is only controlled by the range of the local polynomial.

The temporal trends in the regression are fit with random walk models with a roughness penalty \citep{langbrezger04, gmrfbook, inla-jrssb09} which shrinks the difference between neighbouring observations. This approach allows for smooth, non-parametric function estimation where the smoothness is controlled through a prior which minimises the difference between neighbouring values in the random walk model. These are discussed further in Section \ref{sec:rw}.

\subsection{Gaussian Markov Random Fields}\label{sec:gmrfs}
The Gaussian process \citep{ohagan78} is a popular method for Bayesian non-parametric spatial \citep{bgfs08}, temporal \citep{gptimeseries} and spatio-temporal smoothing \citep{cressiehuang99, foxflu}. Computation of Gaussian processes, particularly on large data sets, is often cumbersome and, in MCMC simulation, requires long burn-in periods to achieve convergence due to the order of the algorithms for operating on dense precision matrices ($n^3$). Computation time can be significantly reduced by using a Gaussian Markov Random Field approximation to the Gaussian process while still remaining quite accurate for well-posed problems \citep{lindgrenJRSSB}. The introduction of a Markovian structure in the Gaussian prior greatly increases the sparsity of the precision matrix, permitting the use of numerical techniques for sparse matrix algebra.

Gaussian Markov Random Fields (GMRFs) are a special class of multivariate Normal distributions where two vertices (in the graph representing the neighbourhood structure) not connected by an edge have a covariance which is identically zero and are thus conditionally independent \citep{gmrfbook}. GMRFs provide a means of fitting latent Gaussian models, a flexible class of models for statistical inference. An example of GMRFs in statistical inference is the Conditional Autoregressive (CAR) model \citep{besag74} whose neighbourhood matrix is defined by some spatial criterion such as common boundaries of regions, a distance threshold or the nearest few neighbours.

The likelihood for the GMRF with $n$ vertices $\vec{\theta}$ having mean $\vec{\mu} \in \mathbb{R}^{n \times 1}$ and precision matrix $\mat{Q} \in \mathbb{R}^{n \times n}$ is
\begin{equation}
 \pi(\vec{\theta}) = \frac{\left| \mathrm{Q} \right|^{1/2}}{(2\pi)^{n/2}} \exp \left( -\frac{1}{2}\left(\vec{\theta} - \vec{\mu}\right)^T \mat{Q} \left(\vec{\theta} - \vec{\mu}\right) \right).
\end{equation}

As the GMRF is a special case of the Normal distribution, the GMRF is a conjugate prior for GMRFs and the more general Multivariate Normal distribution (also known as a Gaussian Random Field or GRF). The formulation of the regression model as a GMRF rather than a more general GRF with dense precision matrices for each latent model term allows fast computation with the Laplace approximation method (see section \ref{sec:comp}).

\subsection{Random walk penalties}\label{sec:rw}
The B-spline smoothness penalty of \citet{bspline96} can be formulated as a random walk penalty prior \citep{langbrezger04}. The prior is a GMRF with zero mean,
\begin{equation}
\pi \left( \vec{\theta} \middle\vert \tau \right) \propto \tau^{(n-k)/2} \exp \left( -\frac{\tau}{2} \vec{\theta}^T \mat{K} \vec{\theta} \right) \equiv \mathcal{MVN} \left( \vec{0}_{n}, \mat{Q} = \tau \mat{K} \right) \label{eq:Bayessmoothtau}
\end{equation}
where $\tau$ has a non-diffuse prior, e.g. $\tau \sim \Gamma (1, 0.00005) \propto \exp\left( -0.00005 \tau \right)$, to ensure that the posterior is proper, $n$ is the number of basis elements and the penalty matrix, $K$, corresponds to a discretised differential operator of order $k$ whose value at the coefficients is to be minimised. \citeauthor{langbrezger04} applied the smoothing penalty to the coefficients of a B-spline basis but the difference penalty prior can be used as a random walk prior on the values of the covariates themselves \citep[see][Chapter 3]{gmrfbook}.

For $\tau = 0$ in Equation \ref{eq:Bayessmoothtau} we have the case where no smoothing has occurred, so the random walk is equivalent in its formulation to an assumption of independence of neighbouring covariate values. As $\tau$ increases, the non-parametric function estimate becomes smoother with the limiting case that the marginal posterior approaches a constant as $\tau \rightarrow \infty$. To make the prior proper, a small value (e.g. $0.00001$) is added to the diagonal elements of $\mat{K}$ in order to ensure that it is invertible without significantly affecting the other properties of $\mat{Q}$.

\begin{figure}[htb]
\centering
\subfloat[]{\includegraphics[width=0.25\linewidth]{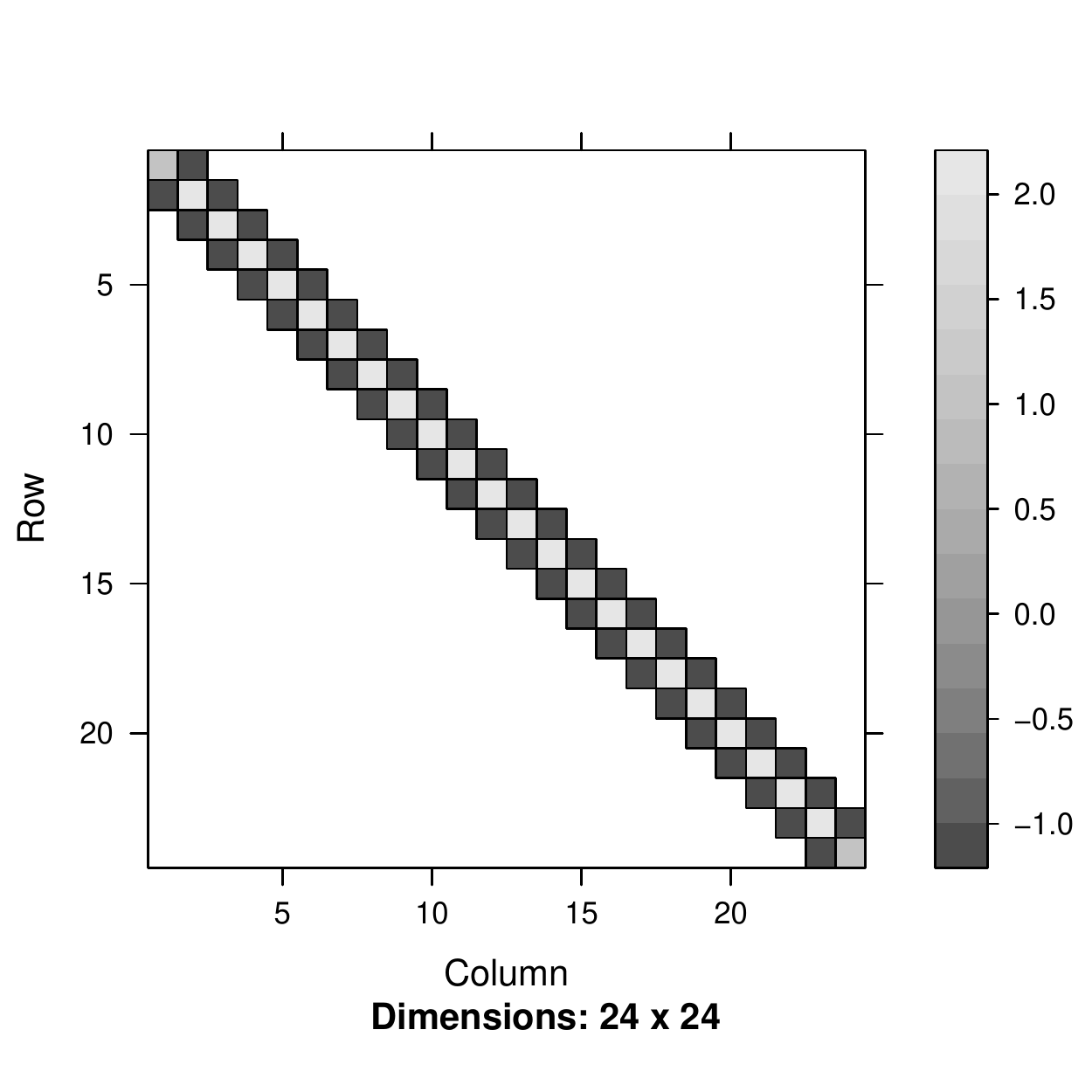}\label{fig:Crw1}}
\subfloat[]{\includegraphics[width=0.25\linewidth]{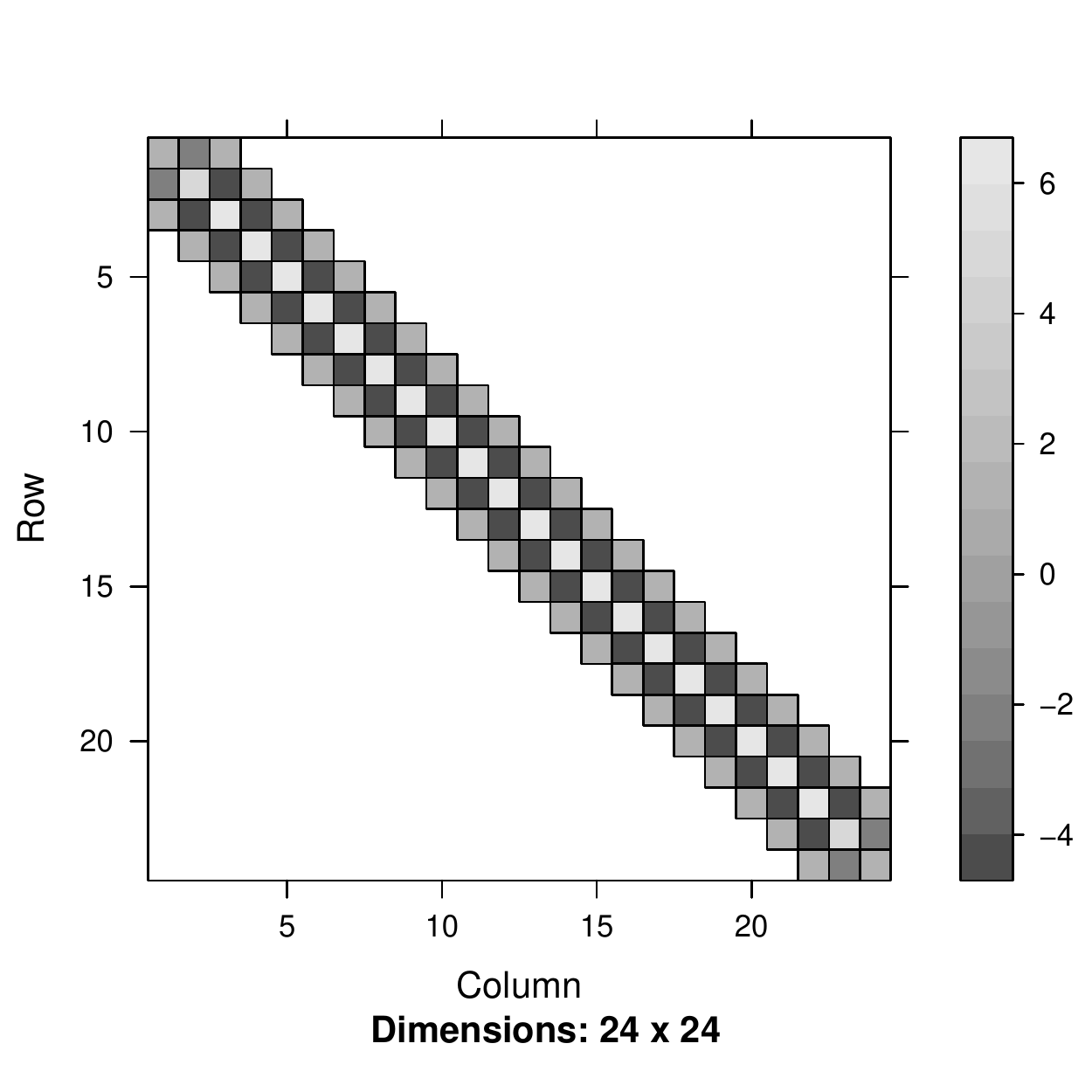}\label{fig:Crw2}}
\subfloat[]{\includegraphics[width=0.25\linewidth]{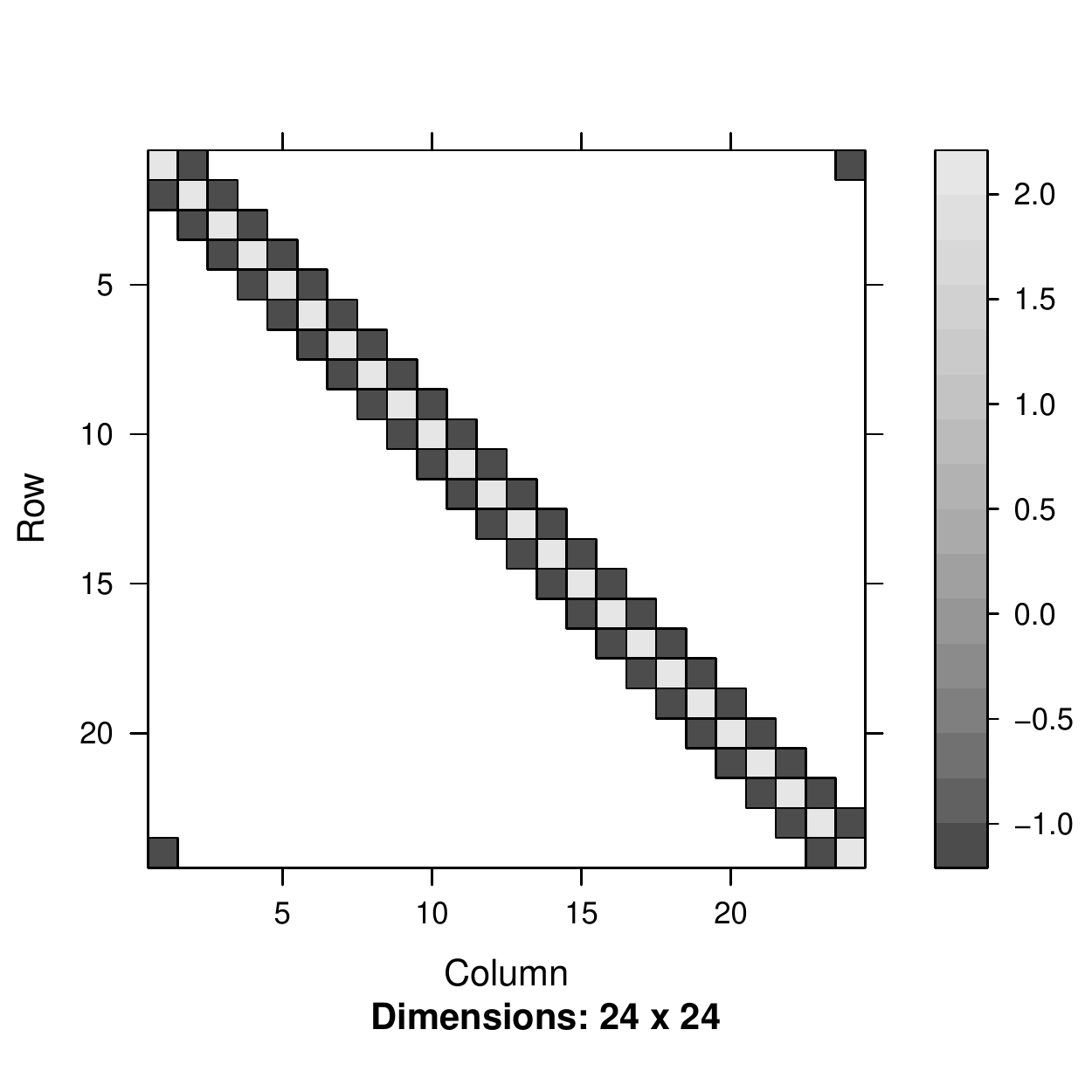}\label{fig:Crw1c}}
\subfloat[]{\includegraphics[width=0.25\linewidth]{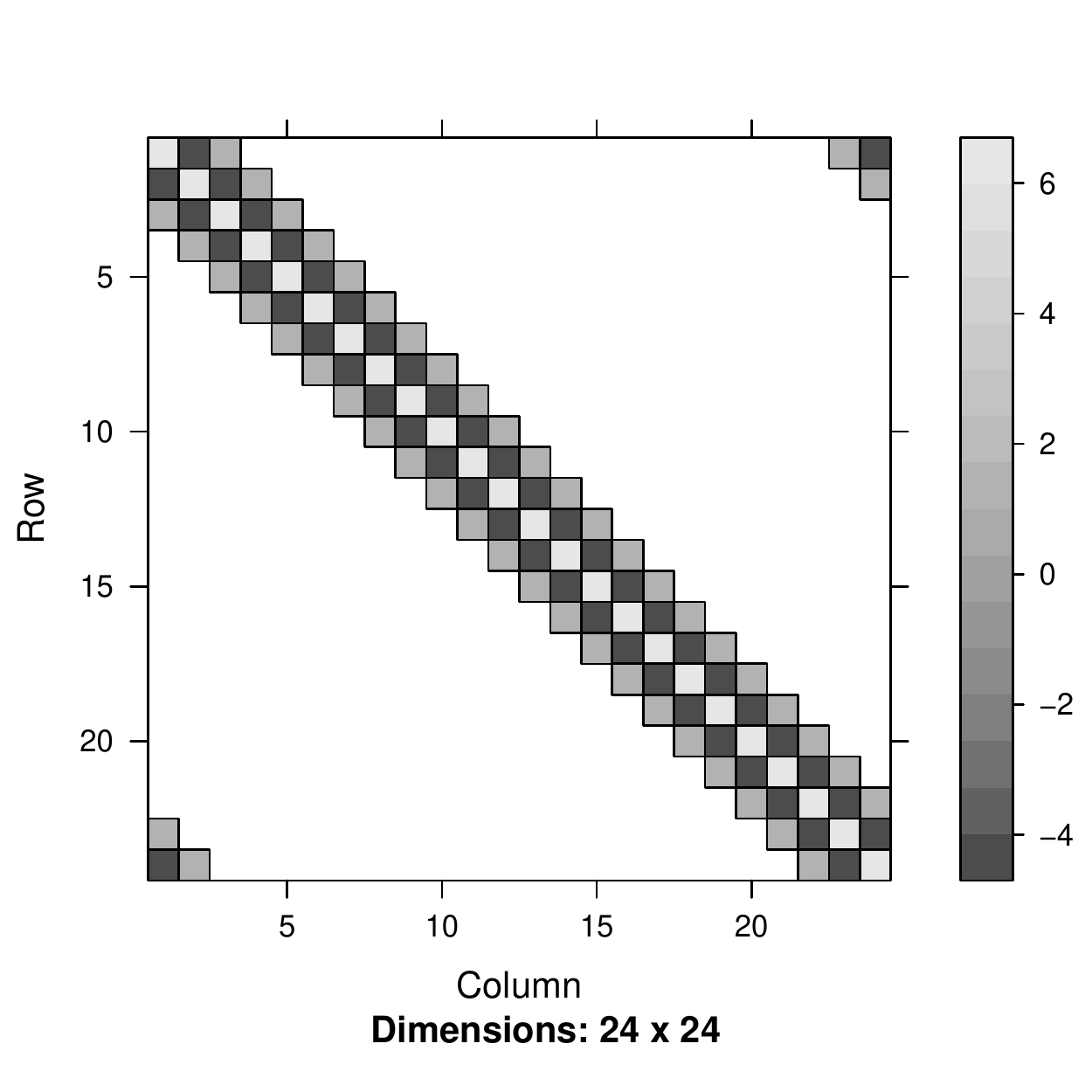}\label{fig:Crw2c}}
\caption{Penalty matrices, $\mat{K}$, for random walk models on equally spaced values on a line. (a) rw1, (b) rw2, (c) rw1c, (d) rw2c.}
\label{fig:Crw}
\end{figure}

It is possible to define periodic versions of the random walk model where the minimum and maximum covariate values are considered neighbours; these cyclic random walks are explored in the next section. First and second order penalty matrices and their cyclic counterparts are shown in Figure \ref{fig:Crw}.

\subsection{Temporal trends}\label{sec:temp}
To fit temporal trends, the hour of the day, day of the week and day of the year are derived from the time stamps of the PNC measurements. Each of these temporal covariates have a periodic nature which must be accounted for in the GMRF's graph and precision matrix.

As there are only seven unique values of day of the week (Sunday, Monday, etc.) there are not enough degrees of freedom to adequately model the weekly trend with a spline, cyclic polynomial or sum of sinusoidal functions. Each day of the week is modelled as a vertex in a GMRF and the weekly trend smoothed by choosing a random walk prior.

To estimate the daily trends, each hour of the day is treated as a vertex in a GMRF prior. The precision matrix for the GMRF is constructed as a Toeplitz circulant difference penalty matrix to account for the periodicity. First and second order random walk precision matrices for such a term can be found in Figures \ref{fig:Crw1c} and \ref{fig:Crw2c}. 

The daily trend may vary with weekday based on local traffic patterns. As such, it is appropriate to construct a covariate which represents the hour of the week, being equal to $24 \left( \mathrm{day of week} - 1 \right) + \mathrm{hrs}$ where $\mathrm{hrs}$ is the hour of the day, from 1 to 24 and $\mathrm{day of week}$ is a numerical coding of the day of the week (with Sunday corresponding to 1).

\begin{figure}[ht]
\centering
\includegraphics[width=0.6\linewidth]{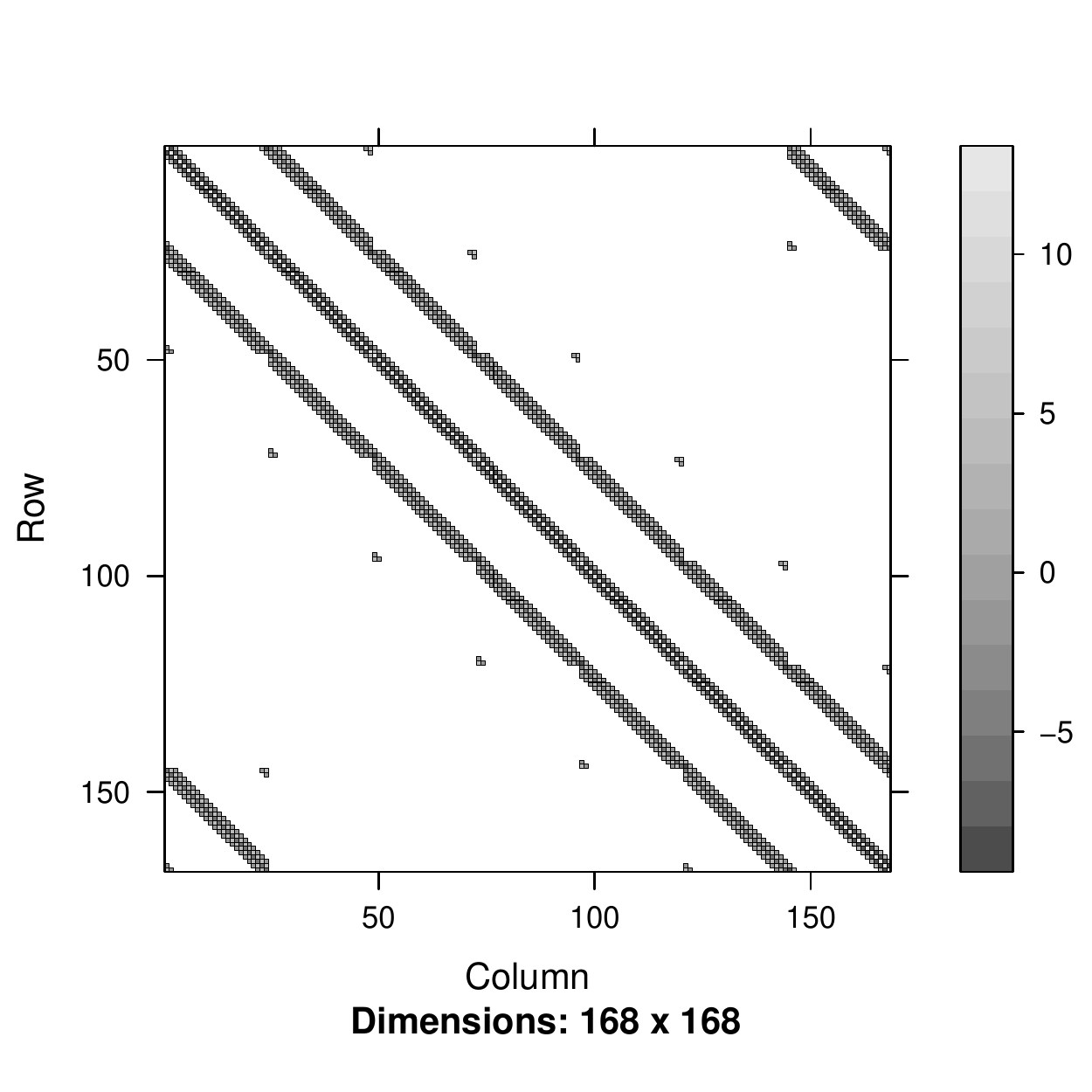}
\caption{Precision matrix for joint model of daily and weekly trends, a Kronecker product of rw2c precision matrices for each of hour of the day and day of the week.}
\label{fig:Qcombblock}
\end{figure}

An appropriate precision matrix for this joint daily and weekly trend is the Kronecker product of the hour of the day difference penalty matrix and the day of the week penalty matrix (Figure \ref{fig:Qcombblock}) \citep{marxeilers2005}. A second order penalty for hour of the day will yield smooth estimates of the daily trend and a first order penalty for day of the week assumes that while there is day to day to day variation the mean level on a Thursday is only related to the concentration on the previous Tuesday through the mean on the Wednesday.

The above precision matrix will be used to model the joint daily and weekly trend at each site as well as a term for all sites, representing the average daily and weekly trend across all of the Metropolitan School District. The hour of the week at each site is treated separately in INLA by replicating the hour of the week as a new column in the data frame for each site. This allows the fitting of two latent models for the same covariate. For the replicated hour of the week covariate for each site, the entries in rows not corresponding to that site are set to ``NA''. This ensures that only observations from that site contribute to the hour of the week term for that site. All of these hour of the week terms are subject to a sum to zero constraint.

By calculating a daily-weekly trend common to all sites, the site-specific daily-weekly trend represents the deviation from the common trend at that site. The daily-weekly trend at each site can be recovered by computing a linear combination of the posterior estimates of the all-sites daily-weekly trend and the site-specific daily-weekly trend. The linear combination is not simply the post hoc sum of the posteriors, as the all-sites and site-specific trends may not be conditionally independent.

The weekly trend at each site can be recovered by forming a linear combination of the 168 elements of the daily-weekly trend at each site, itself a linear combination, such that the average for each day (Sunday, 1, to Saturday, 7) is the weighted sum of the 24 values corresponding to the hours of the week falling on that day in both the all-sites and site-specific daily-weekly trend. Details of the calculation of linear combinations can be found in Appendix \ref{sec:appweetbix}.

\begin{figure}[ht]
\centering
\includegraphics[width=0.8\linewidth]{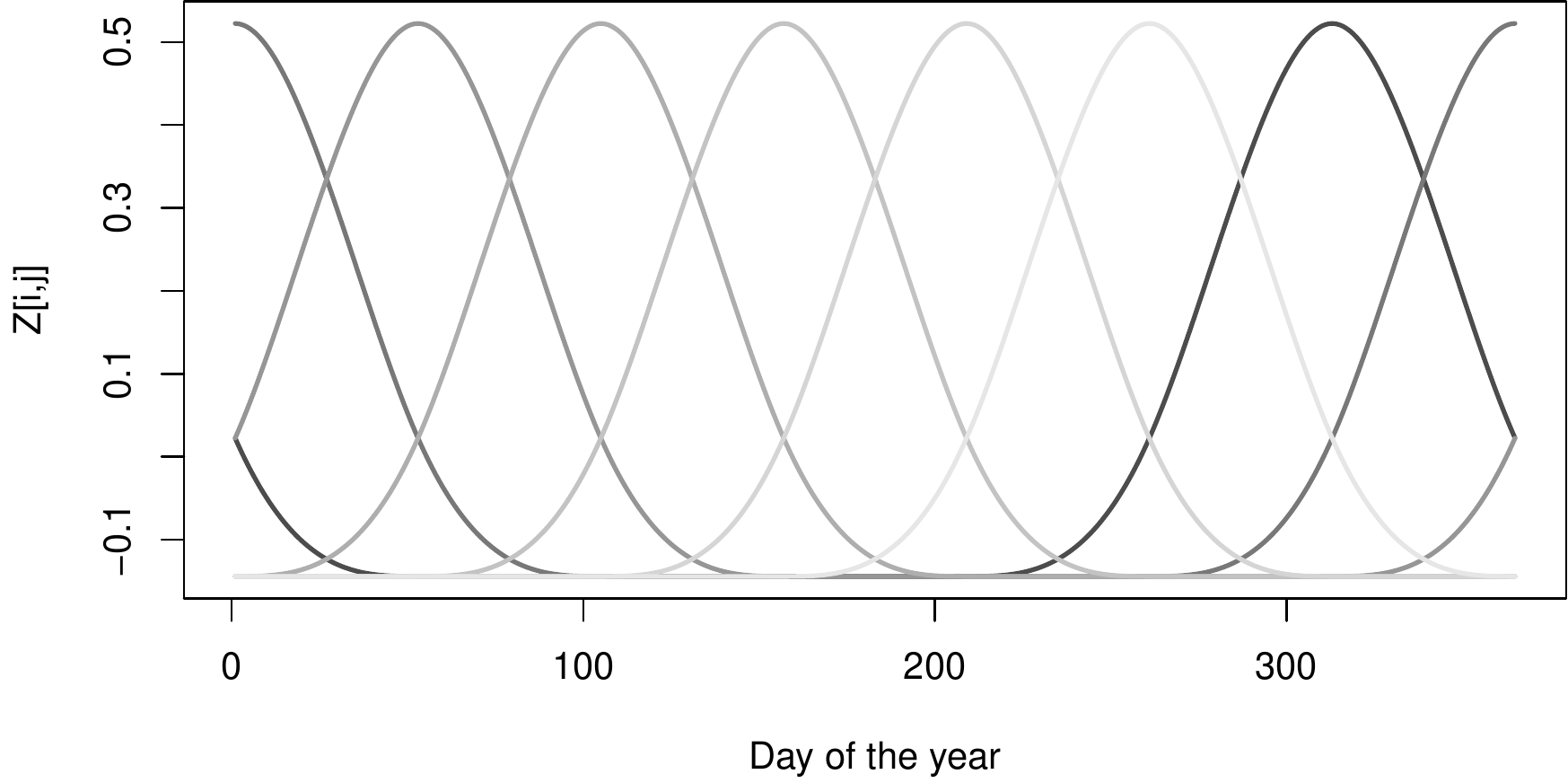}
\caption{Seven zero mean periodic cubic B-spline basis functions for estimating a smooth function of day of the year}
\label{fig:bsp}
\end{figure}

A random walk model (or \texttt{ar1} autoregressive term) could be used to fit the annual trend for all sites, $f_{\textnormal{year}}\left( \mathrm{day of year} \right)$ , but with 366 unique values of day of the year it would result in a non-smooth estimate of the short-term evolution of the daily mean rather than a smoothed annual trend. Instead, the fairly general ``z'' latent model class is used with a custom basis matrix of a cyclic Bezier spline (B-spline) basis  (Figure \ref{fig:bsp}) \citep{deboor78}. The vectors of the basis matrix are then explicitly centred about zero by subtracting the mean of the basis function defined over the values $(1,365)$ from each vector. This ensures that there is no identifiability problem with the full rank of the B-spline basis.

The basis matrix chosen here is constructed with cubic B-splines from a recursive algorithm \citep[see][appendix]{bspline96} defined over a grid of ten knots, yielding seven cubic B-spline basis vectors. The coefficients of the cyclic B-spline basis matrix are constrained to sum to zero and assigned a weakly informative Gaussian prior. The precision parameter for that prior is assigned a Gamma prior
\begin{align*}
\left( \vec{\theta}_Z \pipe \tau_Z \right) \sim & \, \mathcal{MVN}\left(\vec{0}, \tau_Z \mat{I} \right) \\
\tau_Z \sim & \, \Gamma \left(1, 0.00005 \right) \\
       \propto & \, e^{-0.00005\tau_Z}.
\end{align*}

\subsection{Spatial random effect}\label{sec:spde}
The spatial variation in the data may be modelled by assuming that each measurement site has its own mean level. The mean at each site may be related to the means at other sites; in particular, the means of nearby sites may be highly correlated. To model this spatial variation, a structured spatial random effect is specified which takes into account the different mean at each site and the relationship between sites through the use of a stochastic partial differential equation.

The most common approaches to spatial statistical modelling are Kriging \citep{matheron60,spatialstats} and methods based on the Gaussian process \citep{bgcbook, bgfs08}. Kriging and Gaussian process models are equivalent in the case that the observations are assumed to be Gaussian but while these methods are common and simple to apply, they often involve inverting a very dense precision or covariance matrix. Introduction of a Markovian structure to approximate the Gaussian process with a GMRF, as discussed in section \ref{sec:gmrfs}, can improve computational speed without necessarily sacrificing accuracy.

\citet{lindgrenJRSSB} provide a latent model for fitting a continuous spatial random effect in R-INLA. This approach discretises a domain with an unstructured triangular mesh and generates a finite element basis from this mesh. The locations of the vertices in the mesh are labelled and a Mat\'{e}rn class covariance function
\begin{equation}
\mathrm{Cov} \left( \vec{\theta}_i, \vec{\theta}_j \right)  = \frac{1}{\tau 2^{\nu - 1} \Gamma(\nu)}  \left( \kappa \left\Vert \vec{\theta}_i - \vec{\theta}_j \right\Vert  \right)^\nu K_{\nu} \left( \kappa  \left\Vert \vec{\theta}_i - \vec{\theta}_j \right\Vert  \right) \label{eq:matern}
\end{equation}
is used to model the covariance between the values at these vertices. The hyperparameters of the Mat\'{e}rn covariance function are the range parameter, $\sqrt{8}/\kappa$, and precision parameter, $\tau$. The order of the Bessel function of the second kind, $\nu$, controls the differentiability of the resulting posterior surface of the spatial random effect.

The particular Mat\'{e}rn covariance function is the solution to a stochastic partial differential equation (SPDE) which characterises the spatial relationship and is defined over the domain of the measurement locations and mesh locations. The hyperparameters of the Mat\'{e}rn covariance function are computed as part of the posterior GMRF.

The stochastic PDE approach with GMRFs here represents an approximation to a Gaussian process (GP) with a  Mat\'{e}rn covariance function. The GP approach considers the covariance between all pairs of points and results in a very dense precision matrix. The SPDE/GMRF approach significantly increases the sparsity of the precision matrix, admitting the use of computational methods for sparse matrices.

Fitting this latent model returns values at the mesh locations. Linear interpolation within each mesh triangle is performed in order to project the spatial random effect on to a lattice for visualisation.

\subsection{Formulation for R-INLA}
The regression model is a semi-parametric regression involving the use of random walk models and other terms which are represented as GMRFs. In this subsection, the representation of each latent Gaussian model term in the regression equation (\ref{eq:model}) is described and the relevant code provided.

The data frame containing the data to be fit is named \texttt{aq.all}. The following variables are included in the data frame:
\begin{itemize}
    \item \texttt{CPC} -- PNC as recorded by Condensation Particle Counter
    \item \texttt{dayno} -- day of the year (1 to 365)
    \item \texttt{hrofday} -- hour of the day (1 to 24)
    \item \texttt{hrofweek} -- hour of the week (1, Sunday, to 7, Saturday)
    \item \texttt{lat} -- latitude of observation location
    \item \texttt{long} -- longitude of observation location
    \item \texttt{idx} -- spatial random effect index, defined by \texttt{inla.mesh.create()}
\end{itemize}

Code for specific tasks can be found in the appendix.

\subsubsection{Computation}\label{sec:comp}
The R library R-INLA \citep{inlapackage} uses GMRFs to perform approximate inference on the more general Gaussian Random Fields, where the Markov assumption simplifies calculation \citep{inla-jrssb09, lindgrenJRSSB}. The integrated, nested Laplace approximation \citep{skeneINLA90} is computed with a Newton solver where the posterior mean is approximated by the posterior mode ($\vec{\theta}^{\ast}$ such that $\pi(\vec{\theta}^{\ast})$ is a local maximum) and the posterior precision is approximated with the Hessian at the mode ($\mat{Q} \approx -\mat{H} |_{\vec{\theta}^{\ast}}$, where $\mat{H}$ is the Hessian matrix) of the posterior density.

To ensure the convergence of the Newton method solver, a successive series of starting values is obtained by using the Gaussian approximation and Empirical Bayes approach to the INLA computation, starting with an assumption of near-independence in the GMRF and gradually reintroduce the Markovian property. This is done by passing the following arguments to \texttt{inla()}
\begin{verbatim}
control.inla = list(diagonal=1e04,
 strategy="gaussian", int.strategy="eb")
control.mode = list(result=starting.value, restart = TRUE)
\end{verbatim}
The \texttt{diagonal} value is then sequentially reduced by an order of magnitude or two each time, taking the previous solution as \texttt{starting.value}. This sequential reduction gradually reintroduces the Markov assumption, using the previous parameter estimates as a starting point for the Newton solver. Once the starting values have been computed with a low \texttt{diagonal} amount (e.g. \texttt{1e-5}) the solution for inference is computed with Laplace, rather than Gaussian, approximations with
\begin{verbatim}
r1 <- inla(...,
 control.mode=list(result=starting.value, restart=TRUE),
 control.compute=list(dic=T,mlik=T) )
\end{verbatim}
where \texttt{...} here represent the usual \texttt{data=} and \texttt{formula=} options and any additional arguments.

For each observation, R-INLA calculates the posterior mean, median, standard deviation, 95\% credible interval and Kullback-Leibler divergence (KLD) for all parameters and fitted values corresponding to an observation. These values may be used in posterior predictive checks to ensure that the resulting model describes the data in a sensible manner.

\subsubsection{Temporal trends}
The annual trend is defined with a cyclic, cubic B-spline basis of ten knots, centred about zero. This basis matrix is used within the general latent model class, \texttt{z}, which requires the passing of a basis matrix \texttt{Z}.

The joint daily-weekly trend is defined by deriving the new hour of the week variable as in Section \ref{sec:temp}. A latent Gaussian model of the class \texttt{generic0} is set up, with a precision matrix, \texttt{Q.hrofweek}, which is the Kronecker product of the precision matrices for the hour of the day trend (Figure \ref{fig:Crw2c}, $24 \times 24$) and day of the week trend (as in Figure \ref{fig:Crw1c} but of size $7 \times 7$).

To model the joint daily-weekly trend at each site, the hour of the week variable is replicated for as many schools and long-term monitoring sites there are. For each row in \texttt{aq.all}, the values of the replicated hour of the week is set to \texttt{NA} for all replicated columns other than the one corresponding to the site at which that observation was measured. In this way, for each observation, there will be a non-\texttt{NA} value of \texttt{hrofweek}, corresponding to the daily-weekly trend for all sites, and \texttt{hrofweek.i} for \texttt{i} the location index, corresponding to the daily-weekly trend at school/station \texttt{i}.

\subsubsection{Spatial random effect}
The triangular mesh over which the spatial random effect is fit is created by calling \texttt{inla.mesh.create} and passing the longitude and latitude of the observation locations. To ensure that the mesh is not too coarse in the regions with no observations, a maximum edge length of 0.1 degree is set. Once the mesh is defined, the indices for the mesh vertices at the observation locations are added to the data frame.

\subsubsection{Function call}
The precision matrix from the previous section for the joint daily-weekly trend is reused in the latent model specification for the joint daily-weekly trend at each site.
\begin{Verbatim}
model.hrofweek <- CPC ~ 
  f(hrofweek,   model="generic0", 
    Cmatrix=Q.hrofweek, constr=T, diagonal=1e-3) +
  f(hrofweek.1, model="generic0",
    Cmatrix=Q.hrofweek, constr=T, diagonal=1e-3) +
  f(hrofweek.2, model="generic0",
    Cmatrix=Q.hrofweek, constr=T, diagonal=1e-3) + ...
  f(hrofweek.13, model="generic0",
    Cmatrix=Q.hrofweek,constr=T,diagonal=1e-3) +  
  f(idx,model=uptech.spde) +
  f(dayno,model="z",Z=dayno.Z,constr=T,diagonal=1e-3)
\end{Verbatim}

\section{Results and discussion}
Computation was performed on four nodes of QUT's Lyra supercomputer, an SGI Altix XE cluster. The run time for all computation, including the successive obtaining of starting estimates, was one hour.

The regression terms, represented by GMRFs, are all centred around zero to ensure identifiability. The constant term, $\beta_0$, representing the mean (and median) of the log PNC across the domain of the study, is $8.796$ and has a 95\% credible interval of $(8.671, 8.921)$. Inverting the log transformation yields a median of $6608$ particles per cubic centimetre and a 95\% credible interval of $(5831, 7488)$. 

\begin{figure}[ht]
\centering
\includegraphics[width=0.8\linewidth]{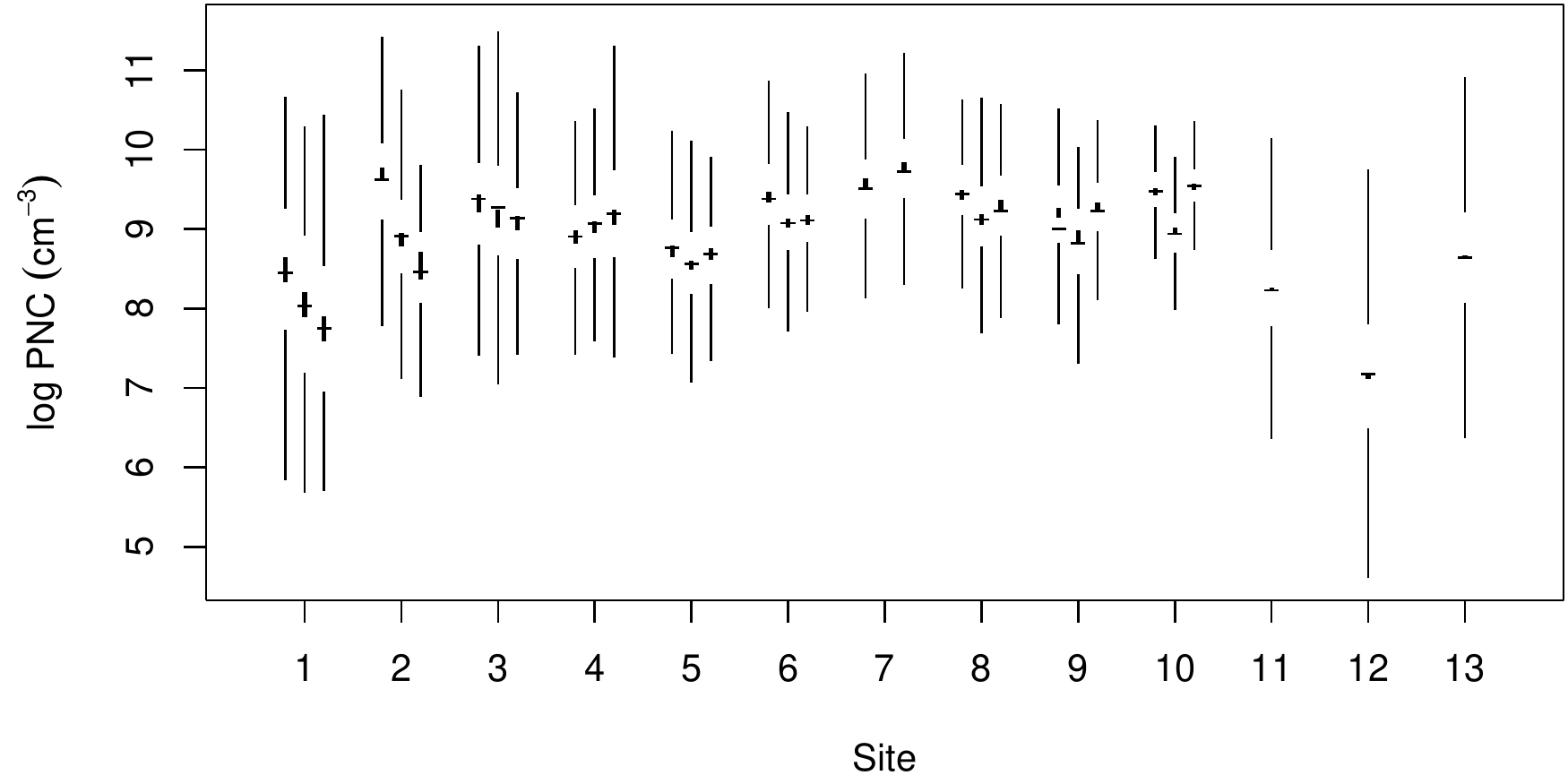}
\caption{Whisker plot of log PNC at each of the 13 measurement sites. Sites 1 to 10 are the ten schools, shown in order of CPC A, B and C within each school. Thin horizontal lines represent the mean; thick vertical lines represent the 95\% confidence interval of the median; thin vertical lines join the first and third quartiles to the most extreme observation which is no further than 1.5 inter-quartile ranges from the median, as in a boxplot.}
\label{fig:sitebox}
\end{figure}

The median, rather than mean, and 95\% CI for the transformed constant are reported because PNC is approximately log-normally distributed (Figure \ref{fig:sitebox}).

\subsection{Spatial random effect}

\begin{figure}[ht]
\centering
\includegraphics[width=0.8\linewidth]{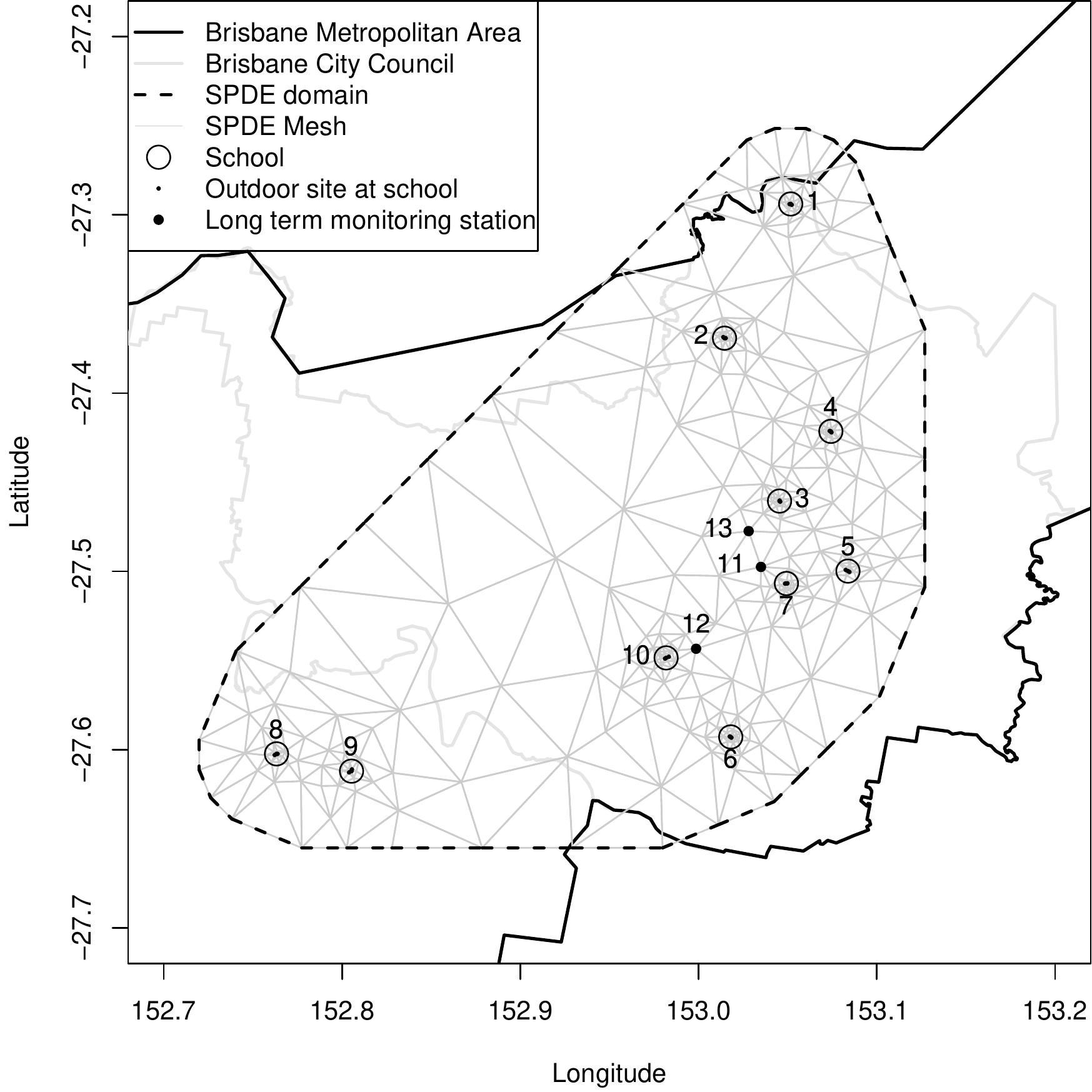}
\caption{Domain of interest for the spatio-temporal model of ultrafine PNC. A triangular mesh is defined within a boundary which contains within it all of the observation locations.}
\label{fig:meshplot}
\end{figure}

\begin{figure}[ht]
\centering
\subfloat[Mean of spatial random effect]{\includegraphics[width=0.4\linewidth]{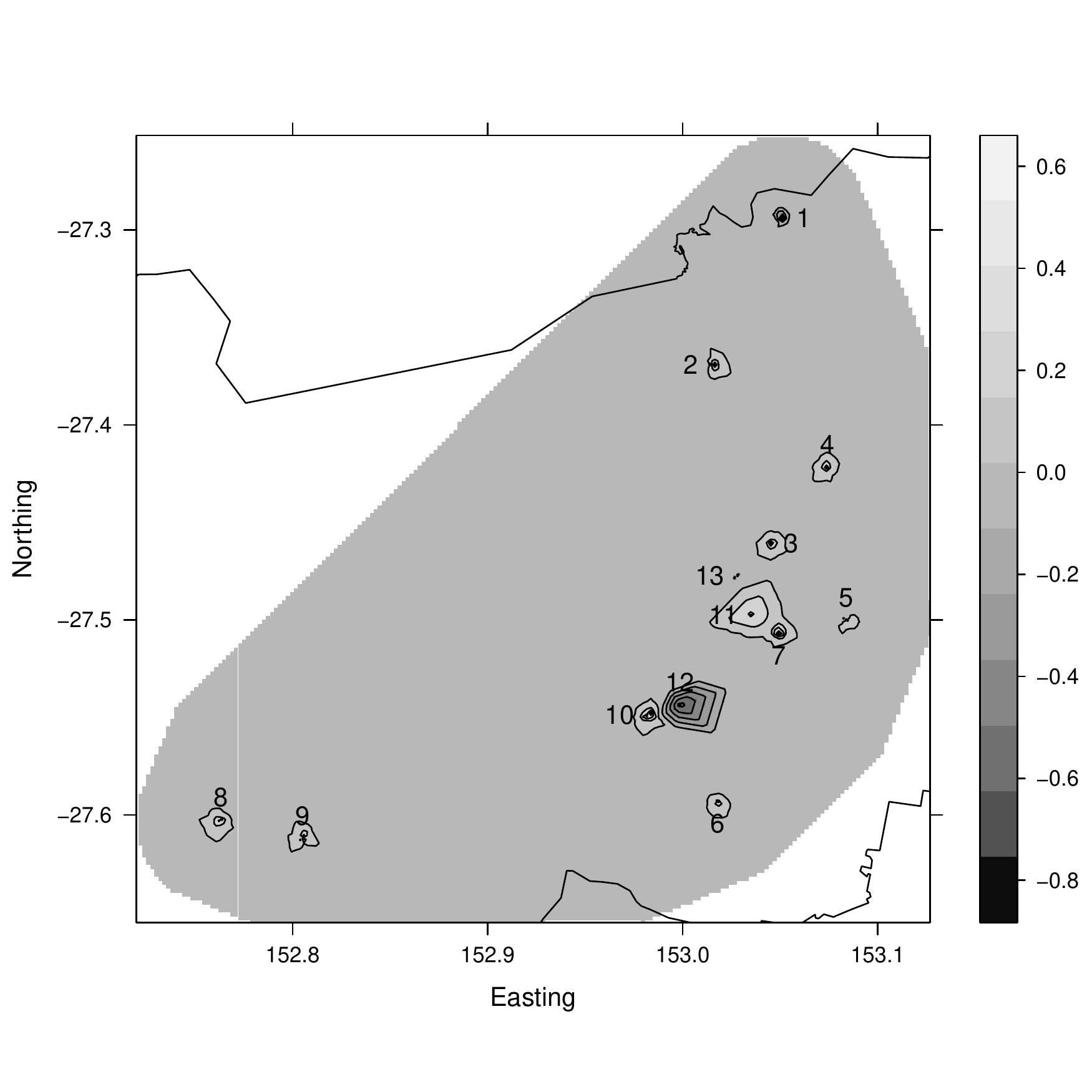}\label{fig:spatmean}} 
\subfloat[Standard deviation of spatial random effect]{\includegraphics[width=0.4\linewidth]{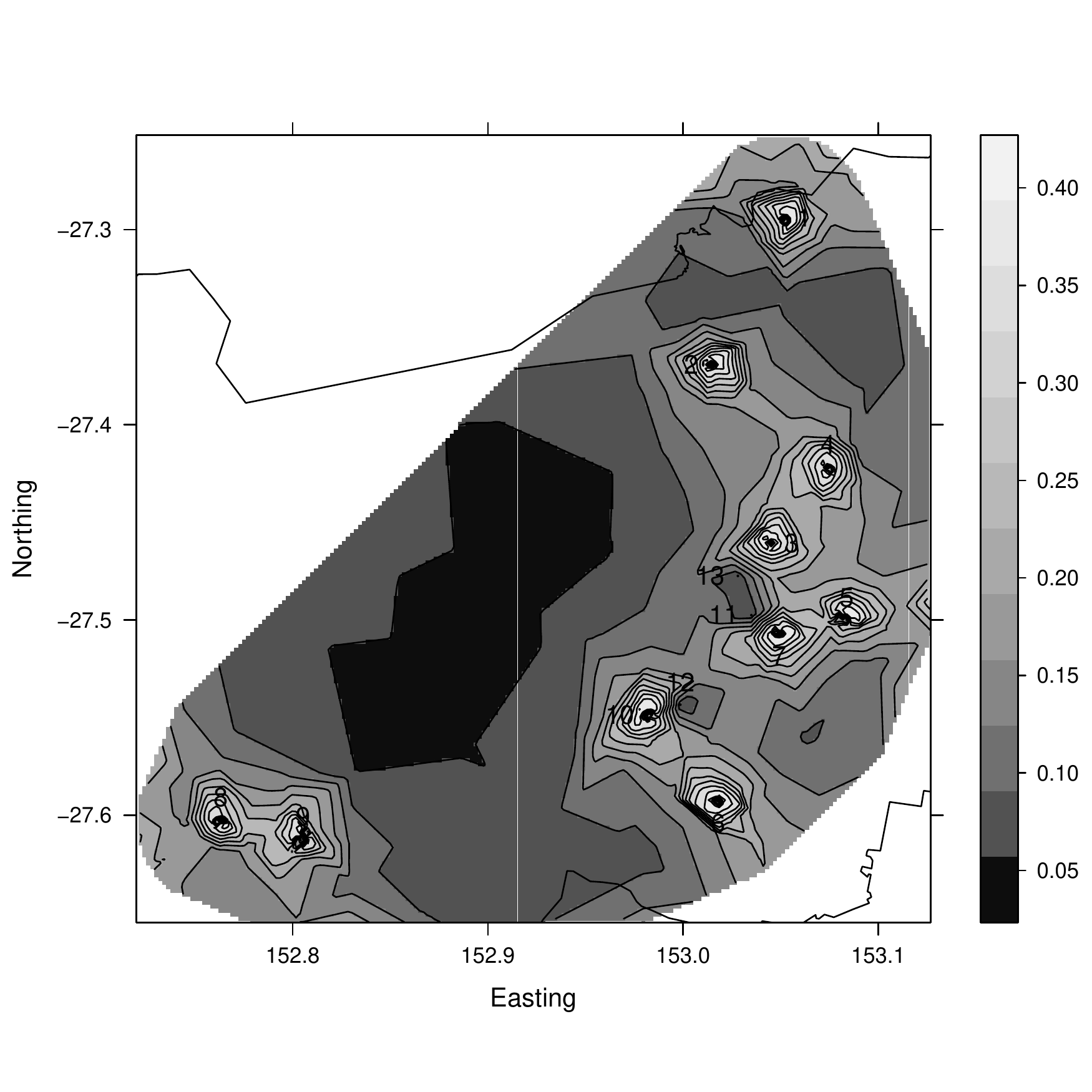}\label{fig:spatsd} }
\caption{Summary statistics of the spatial random effect projected on to a lattice inside the boundary of the SPDE mesh.}
\end{figure}

The spatial random effect (Figure \ref{fig:spatmean}) at the measurement locations  (Figure \ref{fig:meshplot}) indicates that there is a significant amount of spatial variation in the mean level of PNC in the Brisbane Metropolitan Area.

School 1 (indices 1 to 3) is located in a Bayside suburb where the removal of particles by sea breezes is a major contributor to the PNC levels. 

Schools 2 to 10 typically have a spatial random effect which does not contain zero in its 95\% CI, the exception being school 5. Additionally, the spatial effect is such that the first within-school location of each group (corresponding to placement nearest the nearest major road) of three is higher than the other two. The placement at school 4, 7 and 9 has the third CPC nearest the major road, explaining why their estimates generally increase with spatial index.

\begin{figure}[ht]
\centering
\includegraphics[width=0.8\linewidth]{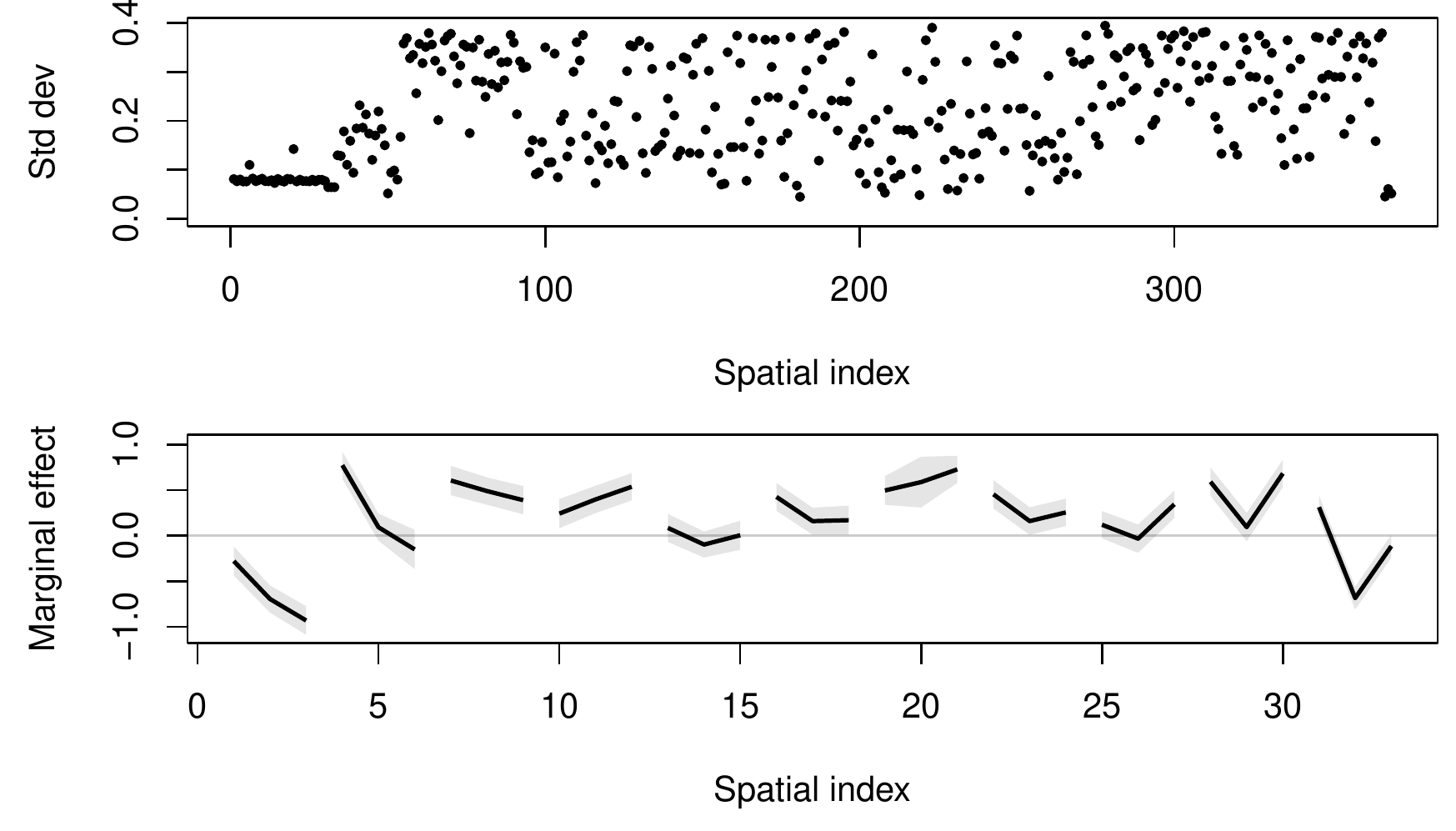}
\caption{Summary statistics of posterior marginals of the spatial random effect. (a) Posterior standard deviation. (b) Posterior mean and 95\% CI at each of the school locations (1--30, grouped three at a time) and long term monitoring stations (31--33).}
\label{fig:spatiall}
\end{figure}

The standard deviation of the spatial random effect (Figure \ref{fig:spatsd}) is smallest at the observation sites (Figure \ref{fig:spatiall}a), particularly at the long term monitoring stations, and is highest a few hundredths of a degree about the observation location. The range of the fitted SPDE (the distance at which spatial correlation is approximately 0.1) is approximately 0.005 degrees (550 m), which covers the school scale spatial variation but not the distance between any two schools or monitoring sites (the minimum value of such a distance is 0.017 degrees, or 1.9 km).

Because the analysis only makes use of data from ten of the 25 UPTECH schools, quite a sparse set of observation locations, the range of the spatial random effect mostly covers the within-school variation rather than the inter-site spatial variation (Figure \ref{fig:spatmean}).

\subsection{Temporal trends}
\begin{figure}[ht]
\centering
\includegraphics[width=\linewidth]{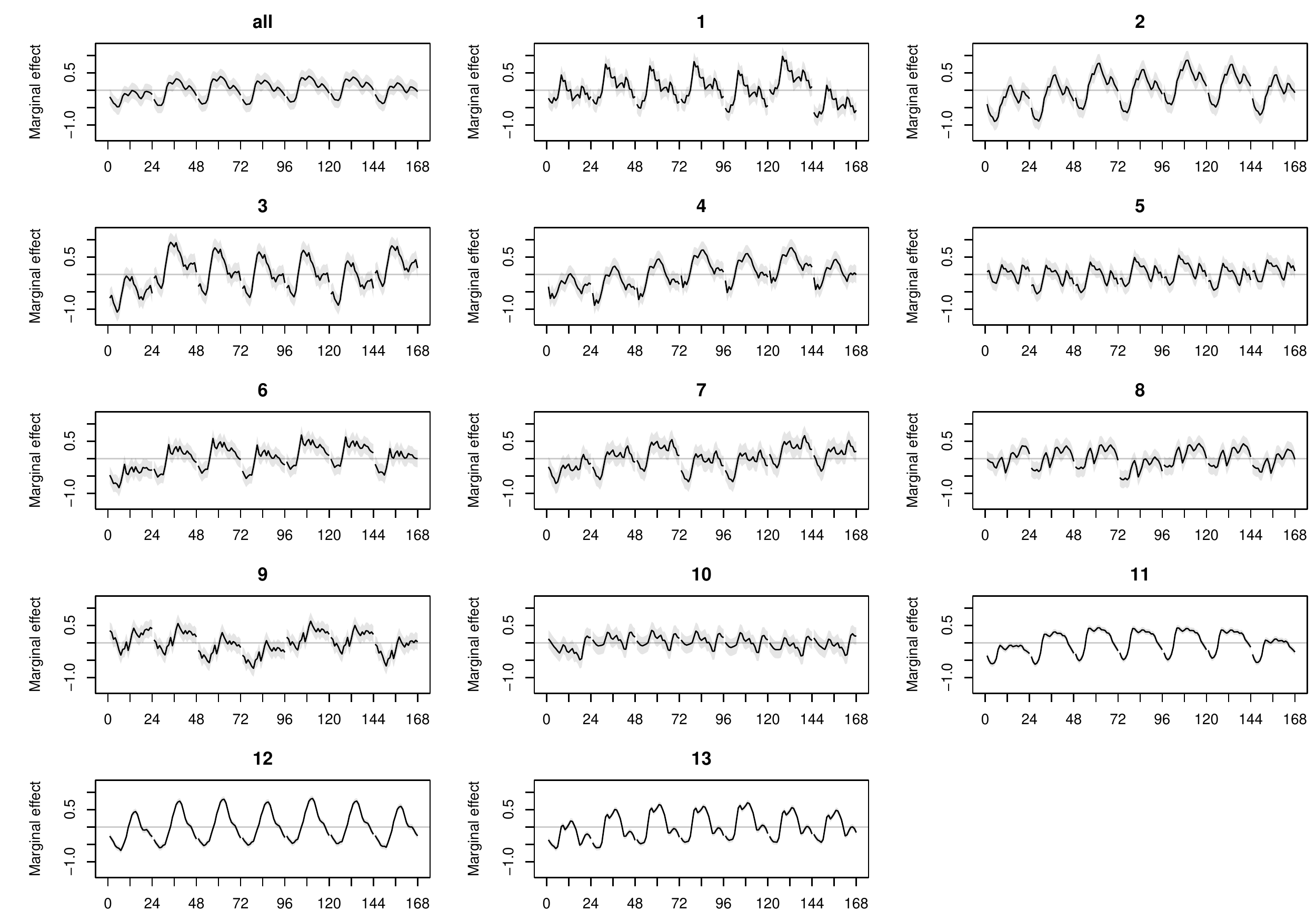}
\caption{Mean temporal trend (and 95\% CI) corresponding to hour of the week across all sites and at each site.}
\label{fig:weetbixweekly}
\end{figure}

After accounting for the mean weekly trend across all sites (Figure \ref{fig:weetbixweekly}, first subplot), the remaining weekly trends at each site differ substantially. The mean daily trend displays peaks during the morning, around midday and a smaller peak in the afternoon. These peaks correspond, respectively, to the morning commute (6-9am), daytime new particle formation events from photochemical reactions \citep{cheung2011, cheungnucleation12} and the afternoon commute (3-6pm). The remaining subplots in Figure \ref{fig:weetbixweekly} show the daily-weekly trend at schools 1 to 10, Woolloongabba (11), Rocklea (12) and QUT (13).

Urban traffic is the source of approximately 80\% of the number of primary particles in Brisbane. The deviation at any site from the all-site trend, especially in the morning and afternoon, are likely due to differences in traffic and wind patterns. Schools 1 and 3 have a morning and evening peak which is more pronounced than the mean daily-weekly trend.

The midday peak coincides with times of maximum insolation, a suggested contributing factor in the formation of new particles during nucleation events \citep{cheungnucleation12}. Nucleation events are typified by a large burst of particles with a diameter smaller than about 25nm, the so-called ``nucleation mode'' \citep{dalmaso07}. The particles in this mode then coagulate, leading to an increase in the count median diameter of the size distribution of particles.

\begin{figure}[ht]
\centering
{\includegraphics[width=0.8\linewidth]{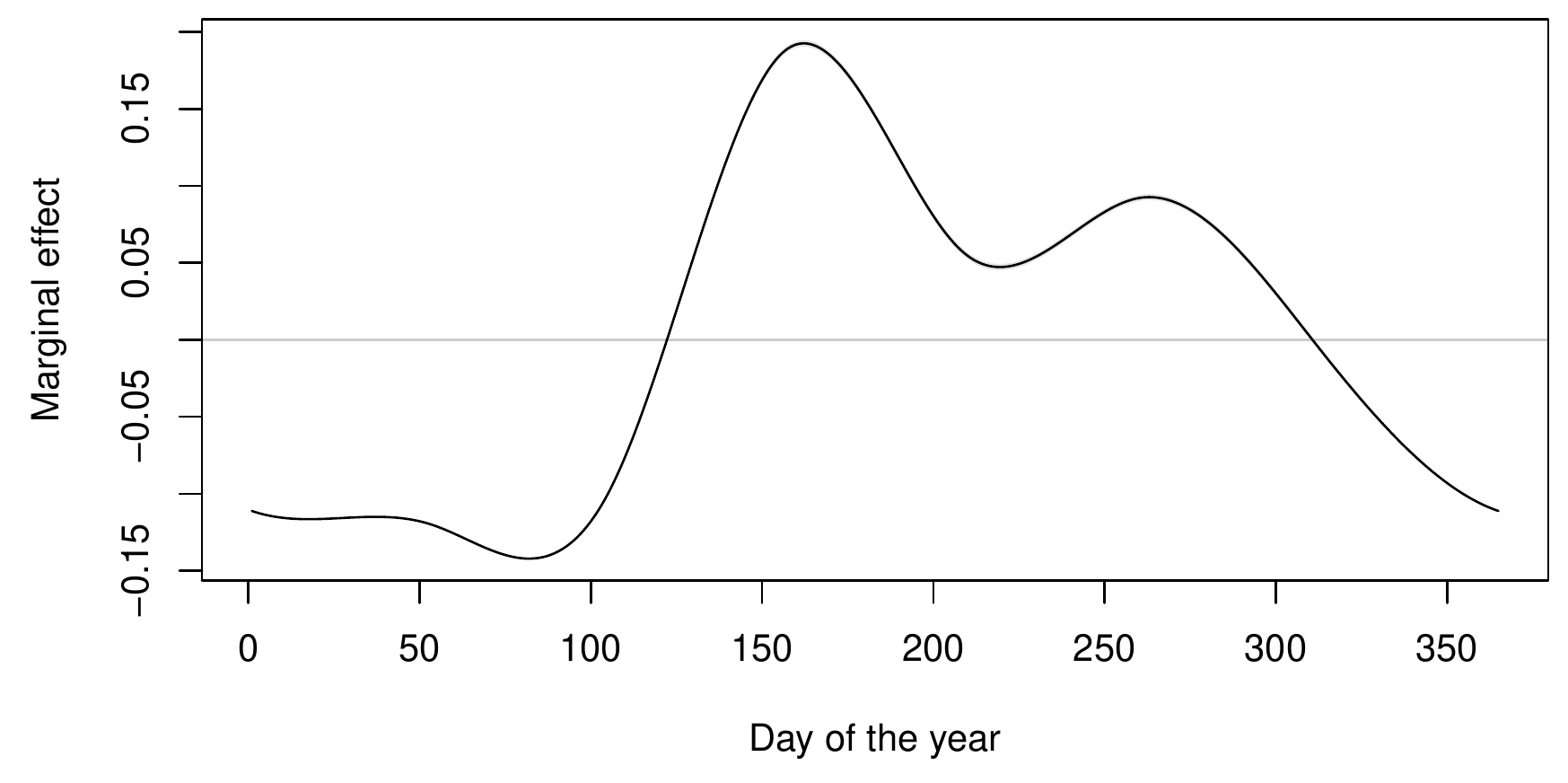}}
\caption{The marginal smooth, annual trend in PNC as a sum of cubic B-splines.}
\label{fig:weetbixannual}
\end{figure}

The midday peak is higher than the all-sites value at schools 2, 3 and 4 and QUT (13) and is lower at 7, 8, 9 and 10. Solar radiation levels vary seasonally, and while the annual trend shows that PNC is highest in winter (Figure \ref{fig:weetbixannual}), when schools 7 to 10 were measured, the annual trend does not explicitly model the effect of sunlight. \citet{jaimenucleation} found that a majority of particle formation events in Brisbane occur in summer and winter, predominantly during daytime hours. While it is postulated that solar radiation plays a role in new particle formation, the frequency of new particle formation events at the Port of Brisbane, an industrial site near the Brisbane Airport in the city's north-east, was found to not be dependent on differences in solar radiation between summer and winter \citep{jaimenucleation}. The prevailing winds during school hours in October to November at Brisbane Airport are east to north-east, so schools 2, 3 and 4 are all downwind of the Brisbane Airport and Port of Brisbane (and have similar prevailing wind patterns). At the QUT site, \citet{cheungnucleation12} found that peaks in PNC were associated with high solar radiation levels and a north-easterly wind blowing from the airport and port. The midday peak at QUT is thus suggestive of transport of new particles formed elsewhere during a nucleation event.

School 7 exhibits an evening peak hour level which is quite high compared to the mean daily-weekly trend. This is also noticeable at QUT (13) and Woolloongabba (11). The daily-weekly trend at school 7 is very similar to the trend at Woolloongabba, with the exception of the evening peak. These three measurement locations are located fairly close together in the corridor of the South-East Motorway, the major road from Brisbane's CBD to the south-east. The mean level of school 7's spatial random effect mean (Figure \ref{fig:spatiall}b) is highest among measurement locations, indicating that the air quality around this school in the evenings (particularly Friday and Saturday) is among the poorest in the study area.

The PNC at school 10 does not vary much in the daily-weekly trend. The variation that is present includes a morning peak around 7-8am and a trough around 6pm. The school is located in the South-Western suburbs of Brisbane and its surroundings include the Brisbane River, a train line, a creek and open green space. The roads surrounding the school are not heavily trafficked in the evening and the nearest motorway is 2km to the South.

Compared to the average mean daily trend, Rocklea (12) has a lower morning peak around 7am. The Rocklea site is located in an open field near a semi-rural industrial site with a motorway to the East and South-East and a major suburban road to the north. Prevailing winds at Rocklea in the morning hours (midnight to 10am) are from south to south-west. That the wind is blowing from the monitoring station towards the motorway and road may explain this reduced PNC level. Conversely, in the afternoon and evening (noon to 8pm), corresponding to the times of high PNC at Rocklea, the wind is blowing from the north-east to south-east, i.e. from the motorway towards the monitoring station.

\begin{figure}[ht]
\centering
\includegraphics[width=\linewidth]{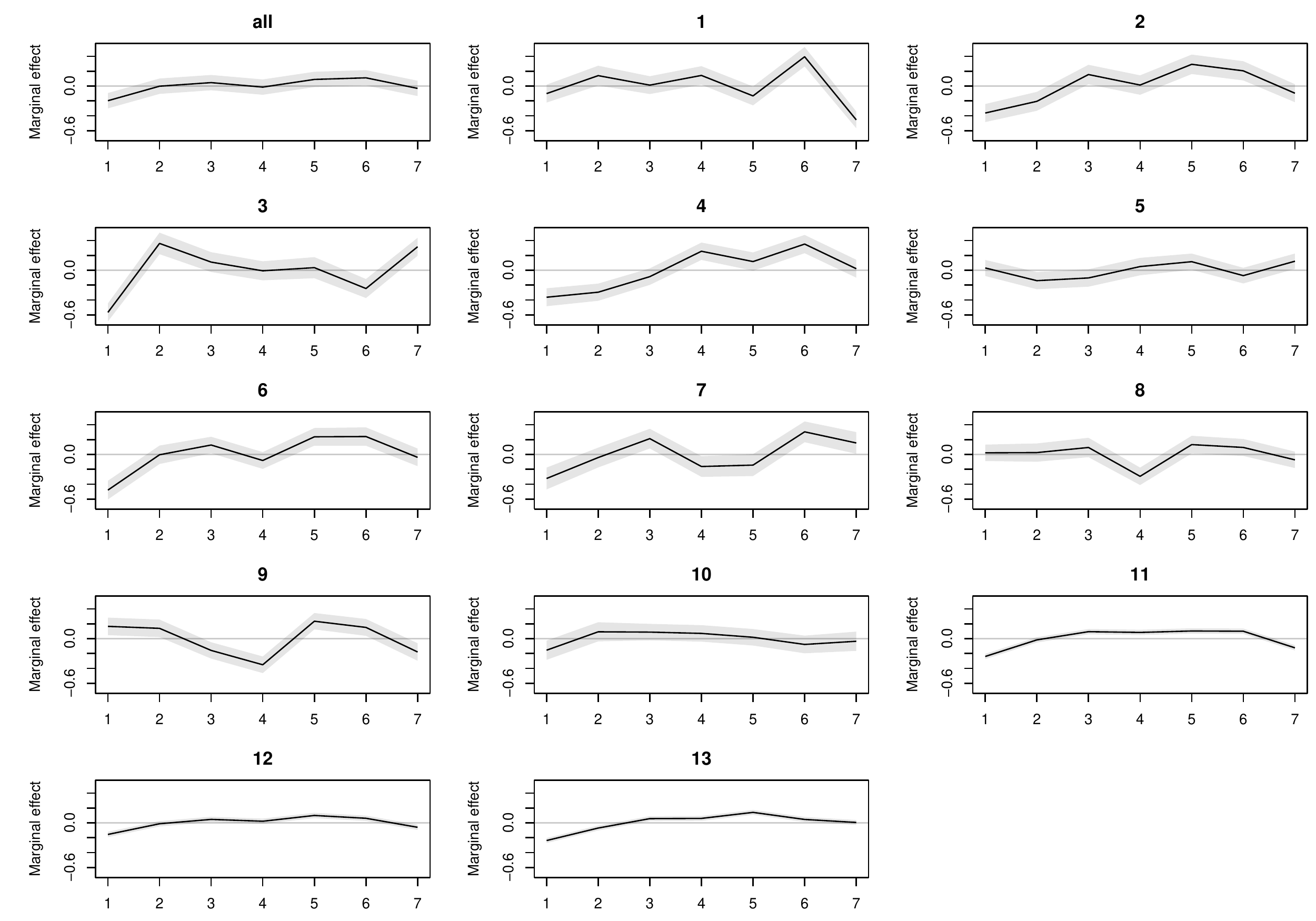}
\caption{Mean temporal trend (and 95\% CI) corresponding to day of the week across all sites and at each site. Derived as a linear combination of marginal effect of hours of the day.}
\label{fig:weeklymar}
\end{figure}

The average for each day of the week is recovered by computing a linear combination of the daily-weekly trend for the 24 observations corresponding to each day (Figure \ref{fig:weeklymar}). The average weekly trend (first subplot) shows that the concentrations in Brisbane tend to be lower on the weekend (days 1 and 7) than on weekdays, with a slightly lower level on Wednesdays. These patterns were examined by \citet{weeklytrends2002} in a longitudinal study of the daily and weekly trends in PNC and its relationship with traffic volume at the QUT site. The day of the week trend at each site is estimated by adding a linear combination of the daily-weekly trend common to all sites to the daily-weekly trend at each site. 

The mean weekly trend across the study area features lower concentrations on the weekend (days 1 and 7). The weekly trend at the long term monitoring stations, sites 11 to 13, contains zero within the 95\% credible interval. The mean weekly trend therefore is representative of the long-term monitoring sites' weekly trend. Sites 1 to 10 (the schools) all have at least one day which does not contain zero in its 95\% credible interval, indicating differences in traffic patterns at each school.

For the daily-weekly trend at each school, there are at most six hourly-averaged observations for each hour of the week (two for each of the three CPCs). As a result, the estimates of the weekly trend for each school are almost a daily average for each of the measurement days. Rather than treating them as independent and identically distributed with a factor term (the \texttt{iid} model in R-INLA), the use of the custom prior precision matrix and linear combination functionality of R-INLA allows the modelling to incorporate information from the entire observation period at each school in the estimate of the weekly trend.

The annual trend (Figure \ref{fig:weetbixannual}) has a very tight credible interval. The long term monitoring stations contribute a lot of information about the annual trend, as all of 2009 was measured by the long term monitoring stations, January 1 to August 31 2010 by the stations at Rocklea and Woolloongabba and September 20 2010 to 16 August 2011 (and beyond) by QUT. In addition to this, the 10 schools in the study so far cover nearly 12 months (although there are gaps). This is equivalent to approximately six years of continuous monitoring at one site. Two peaks are present, in winter and spring.

The range of the effect of the annual trend (0.3) is not as large as that of the daily-weekly trends (0.8). In Brisbane's cool season (May-October), the prevailing long-range winds are from the interior of the continent and transport dry air a low speed. These slow, dry winds prevent the removal of atmospheric pollutants by precipitation \citep{jaime5year}. \citet{jaime5year} conclude that there is no annual trend in Brisbane but it can be seen from the modelling presented here that the annual trend exists but it is not the biggest source of variation in PNC in Brisbane.

A zero-centred cyclic, cubic B-spline was used to model the smooth annual trend. There are other options available for modelling this trend, e.g. by defining a different basis with the ``z'' latent Gaussian model class, which takes a basis matrix as its argument. Other options include using one of R-INLA's pre-compiled model classes (such as a random walk model or autoregressive term or order 1) or by discarding the R-INLA approach altogether. Penalised B-splines \citep{langbrezger04} have been used to model annual trends in ultrafine PNC \citep{finlandpaper} and while R-INLA allows the definition of custom univariate priors (with the \texttt{expression} prior) it is not immediately obvious how to extend this to the multivariate case with a custom basis matrix as in the ``z'' model class.

\subsection{Discussion}
The model of joint daily and weekly trends is more general than a model where they are modelled independently. We see that the daily-weekly trends vary across sites (aim \ref{aim:trend}), implying that to model only the mean temporal trend across the spatial domain would have resulted in ignoring a large amount of information about temporal trends at each site and probably yielded wider credible intervals to represent this.

A spatial random effect for prediction between sites can be estimated with a Gaussian process \citep{bgfs08} instead of the GMRF approximation. This approach uses a Gibbs sampler to perform Bayesian updating of parameter estimates. The slow convergence of MCMC methods in high dimensional data sets, not just for Gaussian process models but for spline models as well \citep{bayesianwinbugs05}, make the Laplace approximation with a Newton solver more appealing. Computation of the dense Gaussian process model with R-INLA would result in not being able to take advantage of methods for approximate inference on sparse matrices. While some simplifications to the spatial model can be made, such as covariance tapering \citep{Kaufman08covariancetapering}, these models may not be sparse enough without using quite a short tapering distance. Recent developments with INLA suggest the ability to combine the Gaussian process approach with INLA \citep{eidsvik12}.

A limitation of INLA at the moment is the inability to model a particular autocorrelation structure in the errors. \citet{finlandpaper} fit a model for ultrafine PNC at one location with a Generalised Additive Model featuring penalised splines and autocorrelated errors, showing that there is a high degree of autocorrelation in PNC even after accounting for the smooth daily trend, weekly and annual trends. INLA does not currently support this degree of structure in the errors, requiring any analysis of the autocorrelation of the errors to be performed post hoc in a second stage of modelling.

\section{Conclusion}
This paper presented a Bayesian semi-parametric spatio-temporal statistical model for data from a split panel design. The model was applied to measurements of particle number concentration in Brisbane, Australia, measured as part of the UPTECH project.

Rather than attempt to directly model the physical system of vehicle exhaust, new particle formation, wind fields and secondary particle reactions, the model estimated the smooth daily-weekly and annual trends and a spatial random effect. The daily-weekly trends were fit with latent Gaussian models with a custom smoothing penalty matrix which combined cyclic smoothing for the hour of the day and day of the week. The trends were modelled simultaneously at all sites and at each individual site to obtain a regional daily-weekly trend and site-specific daily-weekly trends. The continuous spatial random effect was fit by discretising a stochastic PDE over the Brisbane Metropolitan Area.

The uncertainty in the estimates of the spatial and temporal trends was quantified through the 95\% credible interval on parameter estimates, fitted, smooth, semi-parametric functions and their linear combinations. The fitted model explains the spatial and temporal variability inherent in the split panel design rather than fitting an independent model at each site or pooling all data together to fit only the ``all sites'' terms (aims \ref{aim:hier}, \ref{aim:np}).

The all-sites trend exhibited peaks corresponding to the morning and afternoon peak traffic periods as well as a peak during the middle of the day, corresponding to maximum solar radiation and potential new particle formation events. The site-specific trends were seen to be different at each site (aims \ref{aim:hier}, \ref{aim:trend}), with similarities in the midday peak at sites located downwind of the Brisbane Airport and Port of Brisbane according to the prevailing winds at midday, where new particle formation events are suspected to dominate changes in particle number concentration.

In section \ref{sec:spde} the stochastic PDE approach to spatial modelling was applied to model the spatial variation (aim \ref{aim:sites}). While this model term captured the local variation at the school level (aim \ref{aim:loc}) and provided an indication of the mean level for each observation site (schools and long term monitoring stations) the small range of the Mat{\'e}rn covariance function did not provide much in the way of interpolation between sites.

The model was fit to 33 spatial locations, occuring in ten clusters of three and three individual long-term monitoring stations. Whether nucleation is or is not the cause of the midday peaks at QUT and schools 2, 3 and 4 can be investigated by including more schools near the airport and port in the panel design. It is expected that the inclusion of more spatial locations in the panel design will lead to better quality inference about the spatial variation, e.g. whether there is in fact any city-wide smooth spatial trend.

Including meteorological covariates will also account for some of the remaining variability in the data and may affect the range of the SPDE spatial random effect. Inclusion of traffic data, recorded on the largest road adjacent to the schools, will also be used as an explanatory variable. Modelling these covariates may also aid in identifying whether nucleation plays a major role in ultrafine PNC in Brisbane.

The use of R-INLA to fit the model, represented as a GMRF, made the approximate inference both fast and accurate. The R-INLA package provides a number of useful model classes ``out of the box'' but the ability to define custom precision matrices, through the \texttt{z} and \texttt{generic0} model classes and matrix operations such as \texttt{kronecker()} and \texttt{toeplitz()}, made it possible to model the quite complex temporal relationship between observations (aim \ref{aim:np}).

Successively better starting estimates for the full model were obtained in each case by starting with a Gaussian approximation to the GMRF which weakly approximates the Markovian nature of the GMRF and then gradually strengthens the Markovian assumption with each successive run of the model. This was repeated until the estimate was close enough to the full model that the starting values ensured the convergence of the Laplace approximation (aim \ref{aim:conv}).

The use of R-INLA for the modelling outlined in this paper was motivated by the need for fast inference on the spatial and hierarchical temporal trends in the panel design data without making assumptions about the parametric fom of such trends. The stochastic PDE approach to spatial modelling is arguably more elegant than Kriging, the CAR model or the use of geosplines, where the number of unique spatial locations is a limiting factor in the number of elements in the tensor basis. The implementation of custom latent Gaussian models for the complex modelling of temporal trends was straightforward. The R-INLA code, based on GMRFlib \citep{GMRFlib}, is heavily optimised and is able to make use of supercomputer resources (whether a dedicated supercomputer or a multi-core desktop). The combination of these attributes make R-INLA a very powerful engine for modelling spatio-temporal trends in data from a split panel design.

\section*{Acknowledgements}
Sam Clifford wishes to thank: Professors Noel Cressie and Sudipto Banerjee for feedback on early drafts of this paper; Dr Dan Simpson and Professor H\r{a}vard Rue for help with R-INLA and Gaussian Markov Random fields and Mr Ashley Wright from QUT's High Performance Computing team for help with computing.


The authors are grateful to colleagues at QUT's International Laboratory for Air Quality and Health for data collection at the schools and to the Queensland Department of Environment and Resource Management for data collected at Rocklea and Woolloongabba.

This project was funded by Australian Research Council Linkage granst LP0882544 \textit{Quantification of Traffic Generated Nano and Ultrafine Particle Dynamics and Toxicity in Transit Hubs and Transport Corridors} and the Australian Research Council, Department of Transport and Main Roads (DTMR) and Department of Education and Training (DET) through Linkage Grant LP0990134 \textit{The Effects of Nano and Ultrafine Particles from Traffic Emissions on Children's Health}. 


Our particular thanks go to R. Fletcher (DTMR) and A. Monk (DET) for their vision regarding the importance of this work. We would also like to thank all members of the UPTECH project, including Dr G. Marks, Dr P. Robinson, A/Prof Z. Ristovski, A/Prof G. Ayoko, Dr C. He, Dr R. Jayaratne, Dr G. Johnson, Ms W. Ezz, Prof G. Williams, Messrs L. Crilley, M. Mokhtar, R. Laiman, Ms Nitika Mishra, Professor C. Duchaine, Dr H. Salonen, Dr X. Ling, Dr J. Davies, Dr L.-M. Leontjew Toms, Ms F. Fuoco, Dr A. Cortes, Dr B. Toelle, Mr A. Quinones and Ms P. Kidd for their contribution to this work. We also thank Dr M. Falk for his assistance in the classification of the epidemiological studies and his contribution to the discussion of epidemiological study design, as well as Dr F. Fatokun (formerly from ILAQH, QUT) and Dr J. Mej\'{i}a for the initial literature review, Dr D. Keogh (formerly from ILAQH, QUT) for the organisation of the pilot tests, Prof T. Salthammer and Ms J. Bartsch, Fraunhofer WKI, Germany, for VOC pilot testing and analysis, and Ms R. Appleby and Ms C. Labbe for their administrative assistance.

\bibliographystyle{chicago}
\bibliography{../../Bibliography/allpapers}

\appendix

\section{Code for implementing in INLA}
\subsection{Annual trend}
The R code for generating the B-spline basis is given below and is based on the MATLAB code by \citet{bspline96}.
\begin{Verbatim}
bspline <- function(x,K,bdeg,cyclic,xl=min(x),xr=max(x)){
  # x      - a covariate vector
  # K      - the number of knots
  # bdeg   - the degree of the polynomials in the spline
  # cyclic - whether to make the basis periodic
  # xl, xr - minimum and maximum values in the covariate space
  
  x <- as.matrix(x,ncol=1) # reshape to be a column
  
  ndx <- K - bdeg # how many knots will be left at the end?
  
  # as outlined in Eilers and Marx (1996)
  dx <- (xr - xl) / ndx # step size
  t <- xl + dx * (-bdeg:(ndx+bdeg)) # place knots

  # form spline basis of order 0
  T <- (0 * x + 1) %*% t 
  X <- x %*% (0 * t + 1)
  P = (X - T) / dx
  B = (T <= X) & (X < (T + dx)) 
  r = c(2:length(t), 1) # reordering of indices
  
  # recursive updating of basis, increasing degree each step
  for (k in 1:bdeg){
    B = (P * B + (k + 1 - P) * B[ ,r]) / k; 
  }
  
  # only return the first K columns
  B <- B[,1:(ndx+bdeg)]
  
  # convert to periodic basis
  if (cyclic == 1){
    for (i in 1:bdeg){
    B[ ,i] = B[ ,i] + B[ ,K-bdeg+i]    
  }
  # get rid of the bits that are being reused
  B <- B[ , 1:(K-bdeg)]
  }
  return(B)
}


\end{Verbatim}

The annual trend basis matrix is then
\begin{Verbatim}
dayno.Z <- bspline(aq.all$dayno,K=10,bdeg=3,cyclic=1) - 
             mean(bspline(1:365,K=10,bdeg=3,cyclic=1)[,2])
\end{Verbatim}

\subsection{Daily-weekly trend}\label{sec:appweetbix}
Toeplitz circulant matrices which are analogous to the \texttt{rw1} and \texttt{rw2} prior precision matrices can be calculated with 
\begin{Verbatim}
make.Crw1 <- function(n){
 Q <-toeplitz(c(2,-1,rep(0,n-3),-1))
}
make.Crw2 <- function(n){
 Q <- toeplitz(c(6,-4,1,rep(0,n-5),1,-4))
}
\end{Verbatim}

To generate the prior precision matrix for the hour of the week trend, calculate the Kronecker product of a second order periodic random walk for the 24 hours of the day with a first order periodic random walk for the 7 days of the week.

\begin{Verbatim}
hrsbit <- make.Crw2(24)
weekbit <- make.Crw1(7)
Q.hrofweek <- kronecker(weekbit,hrsbit)
\end{Verbatim}

The posterior weekly trend is computed by specifying a linear combination of the posterior estimates of the daily-weekly trend, 24 at a time each with a weight of 1/24.

For example, the linear combination for the weekly trend common to all sites is defined as
\begin{Verbatim}
lc1.hrofweek = inla.make.lincomb(
 hrofweek = c(rep(NA,0*24),rep(1/24,24),rep(NA,168-1*24)))
lc2.hrofweek = inla.make.lincomb(
 hrofweek = c(rep(NA,1*24),rep(1/24,24),rep(NA,168-2*24)))
lc3.hrofweek = inla.make.lincomb(
 hrofweek = c(rep(NA,2*24),rep(1/24,24),rep(NA,168-3*24)))
lc4.hrofweek = inla.make.lincomb(
 hrofweek = c(rep(NA,3*24),rep(1/24,24),rep(NA,168-4*24)))
lc5.hrofweek = inla.make.lincomb(
 hrofweek = c(rep(NA,4*24),rep(1/24,24),rep(NA,168-5*24)))
lc6.hrofweek = inla.make.lincomb(
 hrofweek = c(rep(NA,5*24),rep(1/24,24),rep(NA,168-6*24)))
lc7.hrofweek = inla.make.lincomb(
 hrofweek = c(rep(NA,6*24),rep(1/24,24),rep(NA,168-7*24)))
\end{Verbatim}

In a similar way, the daily-weekly trend at a given site can be recovered by constructing a linear combination of the daily-weekly trend common to all sites and the daily-weekly trend term specific to that site. For example, for site 12 (Rocklea),
\begin{Verbatim}
for (hrofweek in 1:168){
    index <- NA*(1:168)
    index[hrofweek] <- 1
    assign(paste("lc.",hrofweek,".hrofweek.12",sep=""),
     inla.make.lincomb(hrofweek = index,
                    hrofweek.12 = index))
}
\end{Verbatim}

Further details of the calculation of linear combinations can be found at the R-INLA FAQ\footnote{\url{http://www.r-inla.org/faq}}.

\subsection{Spatial random effect}
\begin{Verbatim}
uptech.mesh <- inla.mesh.create(loc=aq.all[,c("long","lat")],
                refine=list(max.edge=0.1))
uptech.spde <- inla.spde.create(uptech.mesh, model="matern")
aq.all$idx <- uptech.mesh$idx$loc
\end{Verbatim}
The function \texttt{inla.spde.create} creates the latent Gaussian model for the stochastic PDE whose solution is the Mat{\'ern} class covariance function (with default order $\nu = 2$).

\end{document}